\documentclass[aps,prb,showpacs,preprintnumbers,amsmath,amssymb,superscriptaddress,twocolumn]{revtex4-1}
\usepackage{amsmath}
\usepackage{graphicx}
\usepackage{hyperref}
\usepackage{xcolor}
\usepackage{float}
\usepackage{ulem}
\usepackage{CJK}

\graphicspath{{./figures/}{./figures_supp/}}

\newcommand{\bs}{\boldsymbol}

\newcommand{\ssec}[1]{\emph{#1} ---}
\newcommand{\midrule}{\hline}
\newcommand{\bottomrule}{\hline\hline}

\begin{document}
\title{Metal-insulator transition in transition metal dichalcogenide heterobilayer: \\
accurate treatment of interaction}
\author{Yubo Yang}
\affiliation{Center for Computational Quantum Physics, Flatiron Institute, New York, NY, 10010, USA}
\author{Miguel Morales}
\affiliation{Center for Computational Quantum Physics, Flatiron Institute, New York, NY, 10010, USA}
\author{Shiwei Zhang}
\affiliation{Center for Computational Quantum Physics, Flatiron Institute, New York, NY, 10010, USA}
\date{\today}
\begin{abstract}
Transition metal dichalcogenide superlattices provide an exciting new platform for exploring and understanding a variety of phases of matter.
The moir\'e continuum Hamiltonian, of two-dimensional
jellium in a modulating potential,
provides a fundamental model for such systems.
Accurate computations with this model are essential for interpreting experimental observations and making predictions for future explorations.
In this work, we combine two complementary quantum Monte Carlo (QMC) methods, phaseless auxiliary field quantum Monte Carlo and fixed-phase diffusion Monte Carlo, to study the ground state of this Hamiltonian.
We observe a metal-insulator transition between a paramagnetic and a $120^\circ$ N\'eel ordered state as the moir\'e potential depth and the interaction strength are varied.
We find significant differences from existing results by Hartree-Fock and exact diagonalization studies.
In addition, we benchmark density-functional theory, and suggest an optimal hybrid functional which best approximates our QMC results.
\end{abstract}
\maketitle

\ssec{Introduction}
Correlated insulators~\cite{Tang2020,Xu2020,Regan2020,Wang2020,Shabani2021,Huang2021} and other interaction-driven electronic states~\cite{Jin2021,Wang2022,Tao2022,Zhao2022} have been realized in moir\'e superstructures created by the interference between two slightly mismatched 2D crystals~\cite{Mak2022}.
Multilayer transition metal dichalcogenide (TMDC) systems
have become one of the focal points of recent experimental~\cite{Foutty2022,Zhao2022} and theoretical~\cite{Davydova2022,Guerci2022} pursuits.
The low-energy quasiparticles in these semiconductor interfaces traverse a smooth potential energy landscape with moir\'e periodicity, because the band edge energy changes with the local geometry and interlayer coupling~\cite{Li2021d,Nieken2022}, which vary on the moir\'e scale.
The long moir\'e wavelength allows their physics to be largely separated from atomistic details~\cite{Padhi2021}. This creates the opportunity to realize tunable systems whose characteristic density variations are on the moir\'e scale. The strong electron-electron interactions coupled with band engineering and other effects have allowed a fascinating array of quantum phases to be realized.

The moir\'e continuum Hamiltonian (MCH)~\cite{Wu2018,Angeli2021}, of two-dimensional electron gas (2DEG) in a periodic external moir\'e potential, is a fundamental model for such systems.
The MCH is directly connected to experiments.
The external moir\'e potential can be obtained by measuring the band edge variation in a scanning tunneling microscopy (STM) experiment~\cite{Shabani2021,Nieken2022}, while the quasiparticle dispersion can be measured in angle-resolved photoemission spectroscopy (ARPES) experiments~\cite{Wilson2017}.
Further, it contains realistic long-range Coulomb interaction between the electrons.
The MCH is also directly connected to \textit{ab initio} calculations.
It can be derived from a large-scale atomistic density functional theory (DFT) calculation~\cite{Jung2014,Wu2018,Zhang2020,Carr2020,Xian2021} by matching the band structure near the Fermi level.

\begin{figure}[htp]
\includegraphics[width=\linewidth]{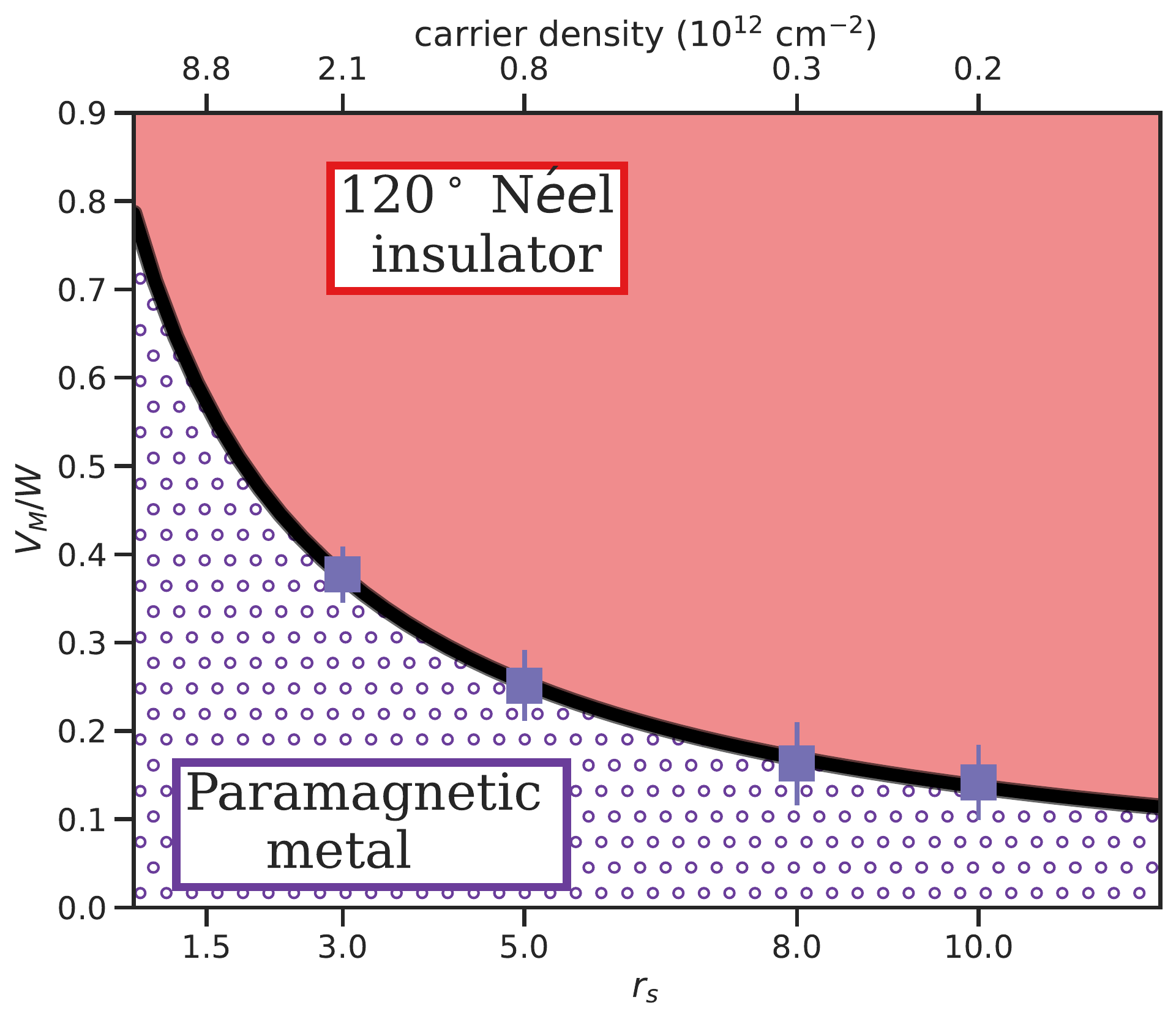}
\caption{Phase diagram of the moir\'e continuum model at half filling.
$V_M/W$ gives the strength of the moir\'e potential,
while $r_s$, the Wigner-Seitz density parameter, is a measure of the interaction strength.
The ground state is
a paramagnetic metal at high density or in a shallow moir\'e potential,
and transitions into a $120^\circ$ N\'eel magnetic insulator with decreasing density or increasing
potential.
The solid black line
identifies a MIT
boundary.
Error bars indicate the estimated systematic uncertainty of the MIT.
The top label maps $r_s$ to hole density in hBN-encapsulated WSe$_2$/MoSe$_2$.
}
\label{fig:dmc-phases}
\end{figure}

Computing the properties of the MCH can provide insight into the physics in 2D materials --- a rich collection has already observed in experiments and undoubtedly much remains to be realized.
In addition, the ability to perform accurate computations for the MCH will allow us to make reliable predictions. Seemingly simple models, such as the uniform electron gas and the Hubbard model, have
provided enormous value in improving our ability to understand and compute much more realistic
and complex materials. The MCH shares this simplicity, and has strong connections to both of these models.
In the deep-moir\'e limit, it downfolds to the Hubbard model on triangular and related lattices~\cite{Wu2018a,Pan2020,Hu2021}.
Although non-local interactions are expected to be important~\cite{Morales-Duran2022,Zhou2022,Gneist2022}.
In the absence of moir\'e, the MCH reduces to the 2DEG, which has long served
as a valuable
model for semiconductor interfaces~\cite{Ando1982,Chaplik1972}.
The inclusion of the moir\'e potential
allows a simple and yet rather realistic modeling
of the environment in 2D TMDC materials.

Much remains to be understood about the properties of the MCH, and little is available in terms of
accurate quantitative information.
In this paper,
we use two complementary many-body QMC methods to explore interesting regimes of the MCH, which involve strong interaction and its delicate interplay with correlation.
No existing theoretical or computational
results
can capture these intricacies with enough reliability to
predict the correlated phases in the model,
which requires accurate treatment of both exchange and correlation effects.
We find a first-order metal-insulator transition (MIT) between a paramagnetic metal and a $120^\circ$ N\'eel insulator.

\ssec{Model}
The MCH, which can be thought of as an effective model for holes in the valence band at the interface of
TMDC systems,
takes the form
\begin{equation} \label{eq:mch}
H =-\dfrac{\hbar^2}{2|m^*|} \sum\limits_i \nabla^2_i - V_M \sum\limits_i \Lambda(\bs{r}_i) +\dfrac{e^2}
{4\pi\epsilon}
\sum\limits_{i<j} \dfrac{1}{\vert \bs{r}_i-\bs{r}_j\vert},
\end{equation}
where $m^*<0$ is the hole effective mass and $\epsilon$ is the permittivity of the dielectric environment.
The parameters $V_M$ and $\phi$ define the depth and shape of the moir\'e potential.
We take
$\Lambda(\bs{r}) =\sum\limits_{j=1}^3 2\cos(\bs{r}\cdot\bs{g}_j+\phi)$,
where $\bs{g}_j$
are three of the smallest non-zero reciprocal lattice
vectors
of the moir\'e unit cell.

Given a filling factor $\nu$, the moir\'e lattice constant $a_M$ defines the Wigner-Seitz radius $a$,
which sets the kinetic and interaction energy scales,
$W\equiv\frac{\hbar^2}{\vert m^*\vert a^2}$ and $U\equiv\frac{e^2}{4\pi\epsilon a}$.
Defining an effective
$a_B^* \equiv \frac{\hbar^2}{\vert m^*\vert} / \frac{e^2}{4\pi\epsilon}$
as length unit, we can express $a$ in reduced units:
$r_s\equiv a/a_B^*=U/W$.
Using $W$ as energy unit, the MCH in eq.~(\ref{eq:mch}) reduces to
\begin{align} \label{eq:mch-rs}
\mathcal{H} = -\frac{1}{2}\sum\limits_i \tilde{\nabla}_i^2 - \lambda\sum\limits_i \tilde{\Lambda}(\tilde{\bs{r}}_i) + r_s\sum\limits_{i<j} \dfrac{1}{\vert \tilde{\bs{r}}_i-\tilde{\bs{r}}_j\vert}\,,
\end{align}
where
all lengths are scaled by $a$: $\tilde{\bs{r}} \equiv \bs{r}/a$, $\tilde{\bs{g}}\equiv \bs{g}a$ ($\tilde{\Lambda}$ contains $\tilde{\bs{g}}$).
The two parameters, $\lambda\equiv V_M/W$ and $r_s$, fully specify
the system. To connect with experiments, $m^*$ and $\epsilon$ are needed.

We consider $m^*=-0.35$,
$\epsilon/\epsilon_0=4.5$,
$\phi=26^\circ$, and half filling $\nu=1$ as inspired by studies of hexagonal boron nitride (hBN)-encapsulated WSe$_2$/MoSe$_2$ ~\cite{Wu2018a,Zhang2020,Shabani2021,Hu2021,Morales-Duran2021}.
Given these choices,
$a_M=10$ nm corresponds to $r_s = 7.7$.

The actual experimental setup is more complicated than the MCH~\cite{Li2021a,Ghiotto2021,Tang2022} of eq.~(\ref{eq:mch}).
Atomic reconstruction~\cite{Yoo2019,Halbertal2021,Li2022}, gate screening~\cite{Spivak2004,Tang2023}, and disorder~\cite{Tan2022,Ahn2022,Kim2023} all have the potential to make the physics of the device qualitatively different.
However, the MCH Hamiltonian captures essential features which drive
much of the interesting physics in 2D TMDC materials.
The interaction terms in the MCH can be modified to bring the model closer to experiment.
Further, the addition of a spin-orbit term can facilitate the modeling of Janus TMDC bilayers~\cite{Angeli2022}.

\begin{figure}[ht]
\includegraphics[width=\linewidth]{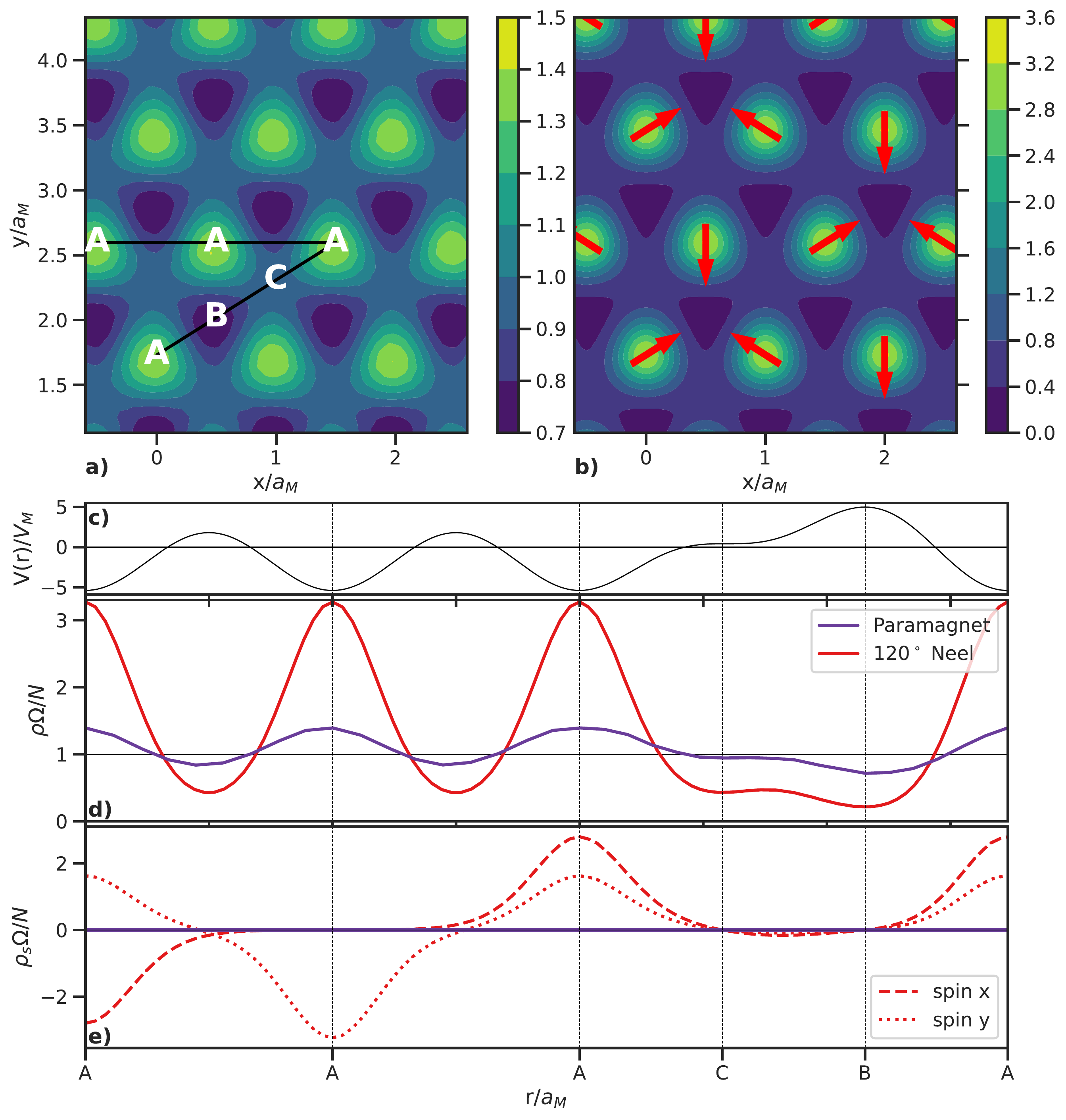}
\caption{Spin and charge densities of representative \textbf{a)} metal and \textbf{b)} insulator
phases (at $r_s=3$ and with $V_M/W=0.2$ and $0.6$, respectively).
The linecuts are drawn for the path shown as a black line in \textbf{a)}.
Panel \textbf{c)} shows the moir\'e potential, while \textbf{d)} and \textbf{e)} show the charge and spin (in 120 N\'eel phase only) densities, respectively.
}
\label{fig:sdens}
\end{figure}

\ssec{Methods}
We apply two QMC methods, diffusion Monte Carlo (DMC) \cite{Foulkes2001} and auxiliary field quantum Monte Carlo (AFQMC) ~\cite{Zhang2003,AFQMC-review-correlated,Hao2021} to study the ground state of the MCH in Eq.~(\ref{eq:mch-rs}).
They
are among the most accurate many-body
methods for strongly correlated systems~\cite{LeBlanc2015,Motta2017,Motta2020,Williams2020}.
One of the major challenges in reliably characterizing the properties of
a system such as the MCH is to maintain accuracy in realistic hamiltonians, which contain long-range Coulomb interactions, and still approach the thermodynamic limit.
The QMC methods we employ allow us to achieve these objectives.

We use the non-collinear spin implementation~\cite{Melton2016,Melton2016a,Melton2017,Melton2019} of fixed-phase DMC (FP-DMC) \cite{Ortiz1993}, and the GPU-accelerated phaseless-AFQMC (ph-AFQMC)~\cite{Zhang2003,Malone2019,Malone2020}.
FP-DMC is variational and works directly in the complete-basis-set limit.
We use it to locate the MIT boundary by comparing the total energies of metallic and insulating states.
Properties including the spin and charge densities and momentum distributions are computed by AFQMC using the mixed estimator\cite{Purwanto2004}.
They are cross-checked with DMC calculations where possible, with consistent results between the two methods.
See Supplemental Material [url] for details, which include Ref.~\cite{Chiesa2006}.

Effective single-particle theories such as Hartree-Fock (HF) and DFT replace the many-body interaction term with an effective single-particle potential.
In this work, we also benchmark their reliability against our QMC results.
One goal of this effort is to identify the best independent-particle approach for 2D TMDC systems, which will greatly help initial screening of basic properties using relatively inexpensive and quick computations to support the fast-growing experimental effort.
It is important to emphasize that these benchmarks are only a first step, however, since the performance will vary as we vary the system parameters (including, among others, $\nu$ and $\phi$).
We perform DFT calculations via the local density approximation (LDA)~\cite{Perdew1992} as well as hybrid functionals~\cite{Becke1993}.

Our QMC calculations are typically performed in $36$- and $144$-electron systems with $4\times4$ and $2\times2$ twist-averaged boundary condition, respectively.
Structure factor based finite-size correction~\cite{Holzmann2009,Holzmann2016-fsc,Yang2020-CP} and grand-canonical twists~\cite{Lin-TABC} are used to obtained the momentum distribution in the metallic phase.
All DMC calculations use a fictitious spin mass of $500$ a.u. to sample spins.
In FP-DMC, we use a Slater-Jastrow wavefunction ansatz, which is optimized with variational Monte Carlo.
The Jastrow contains short-range two-body correlations, represented by B-splines.
Our ph-AFQMC calculations are performed using a Kohn-Sham orbital basis.
We use single Slater determinant trial wavefunctions generated using either HF or LDA.
The lowest-energy trial is used to calculate QMC properties. In the paramagnetic metal phase, the LDA trial is chosen, otherwise the insulating HF trial has lower energy.
The QMC calculations are carried out using QMCPACK 3.15.9~\cite{Kim2018,Kent2020} with appropriate 2D modifications.
We perform HF and DFT calculations using quantum espresso (QE) 7.1~\cite{Giannozzi2009,Enkovaara2017}, modified to perform 2D calculations. We use the 2D LDA functional from libxc 5.1.7~\cite{Marques2012,Lehtola2018}, which is based on DMC data obtained by C. Attacalite \textit{et al.}~\cite{Attaccalite2002}.

\begin{figure}[ht]
\includegraphics[width=\linewidth]{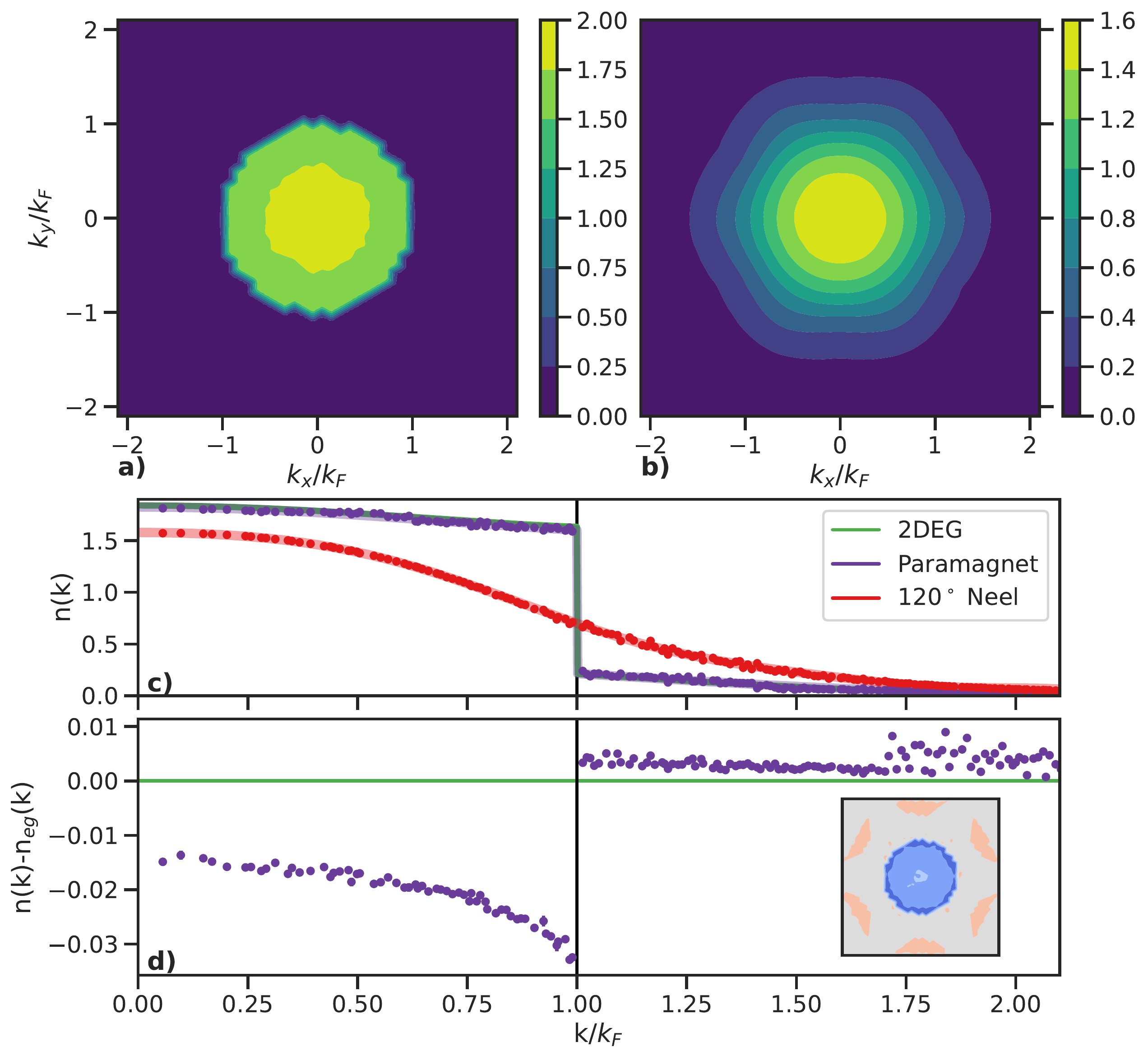}
\caption{Momentum distribution of the same representative systems as in Fig.~\ref{fig:sdens}.
Panels \textbf{a)} and \textbf{b)} show $n({\mathbf k})$ for the metallic and insulating phases, respectively.
Panel \textbf{c)} plots $n(\vert \bs{k} \vert)$ for both systems,
along with that of the 2DEG
for reference.
The metallic system is barely discernible from the 2DEG, both with a discontinuity at $k=k_F$.
Panel \textbf{d)} shows the difference between them with a magnified view.
Secondary Fermi surfaces
are present in the metallic phase, as seen in the inset.
}
\label{fig:nofk}
\end{figure}

\begin{figure*}[ht]
\begin{minipage}{0.32\textwidth}
\includegraphics[width=\linewidth]{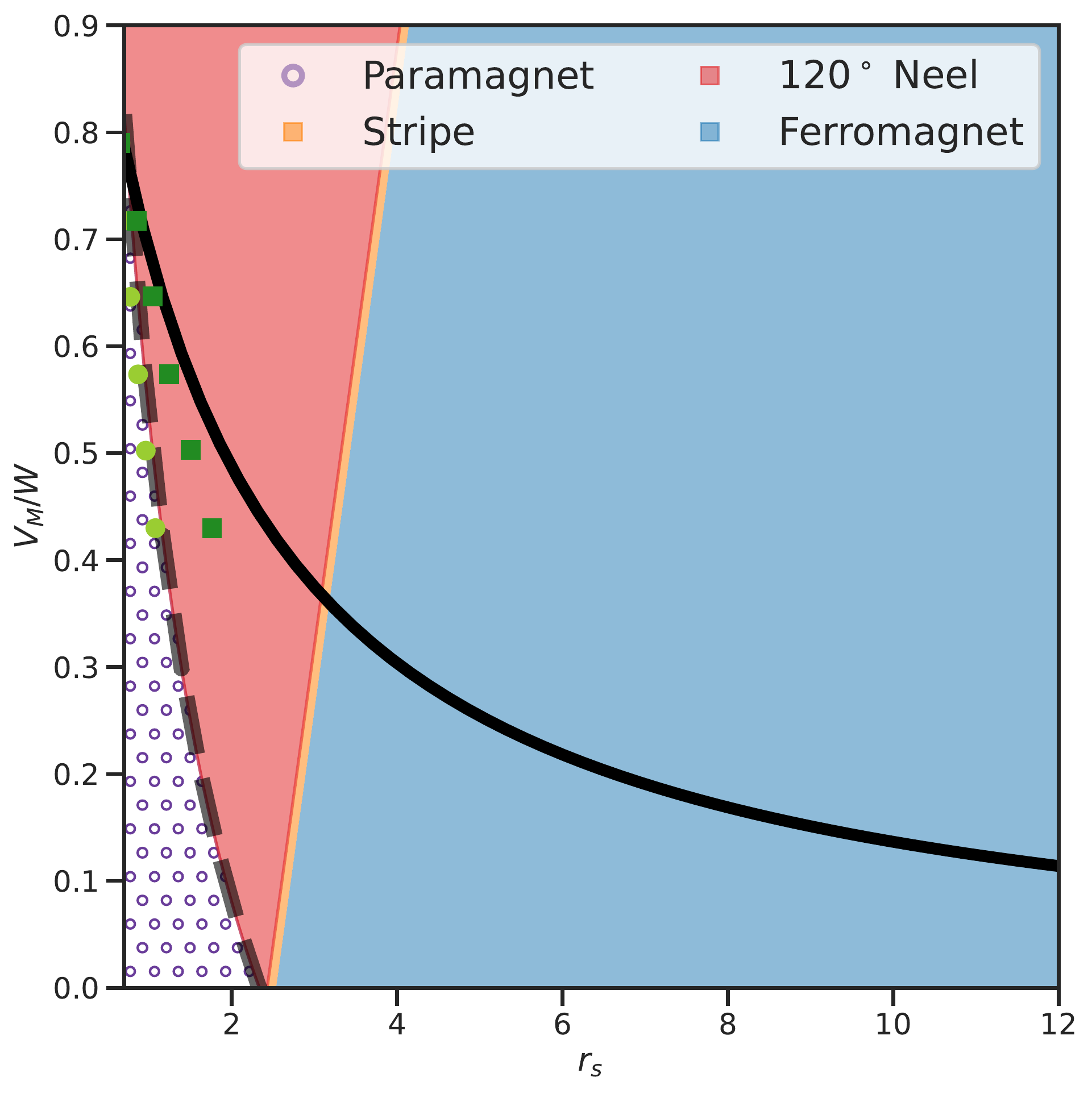}
(a) Hartree-Fock
and exact diagonalization
\end{minipage}
\begin{minipage}{0.32\textwidth}
\includegraphics[width=\linewidth]{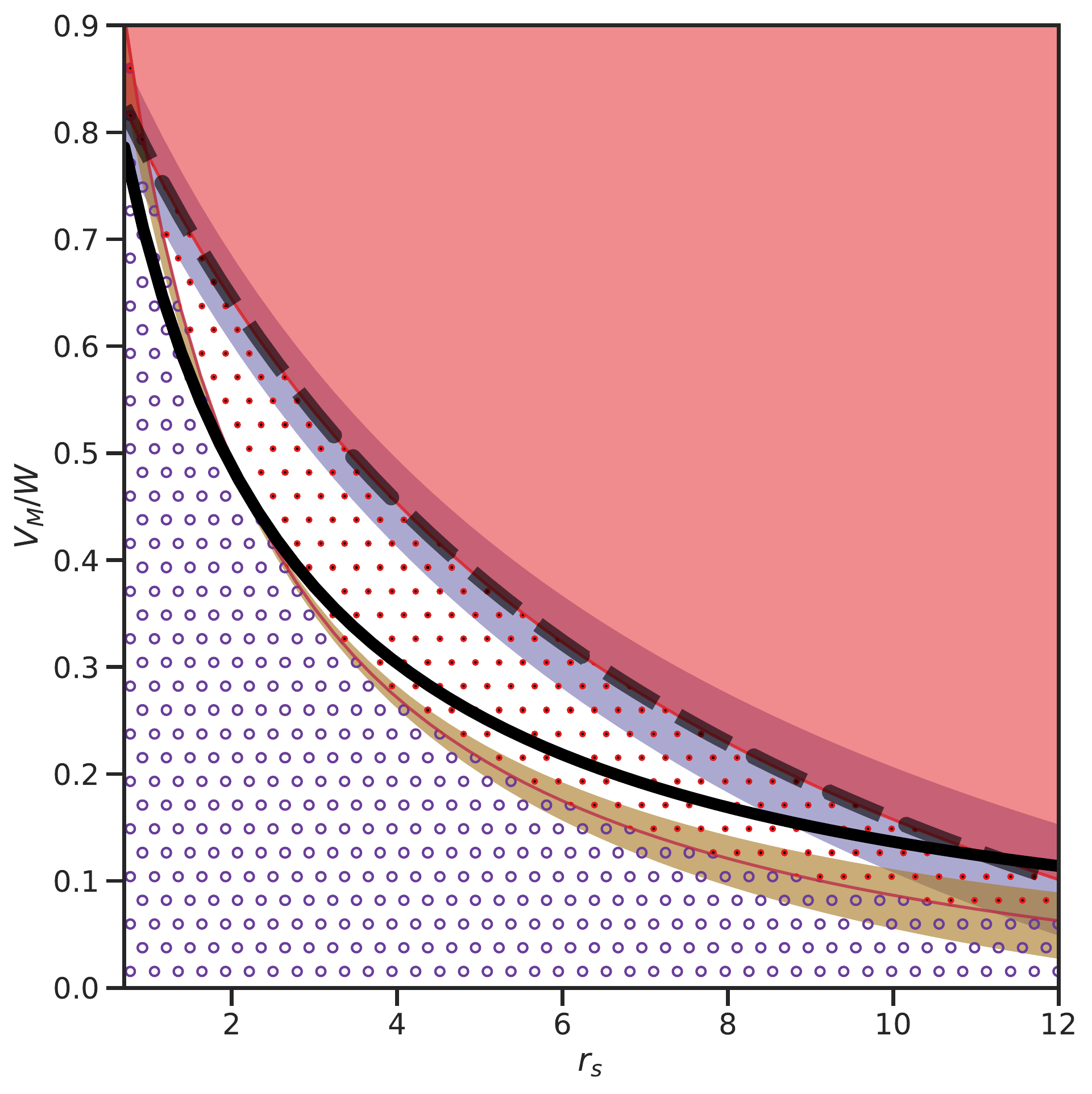}
(b) LDA
\end{minipage}
\begin{minipage}{0.32\textwidth}
\includegraphics[width=\linewidth]{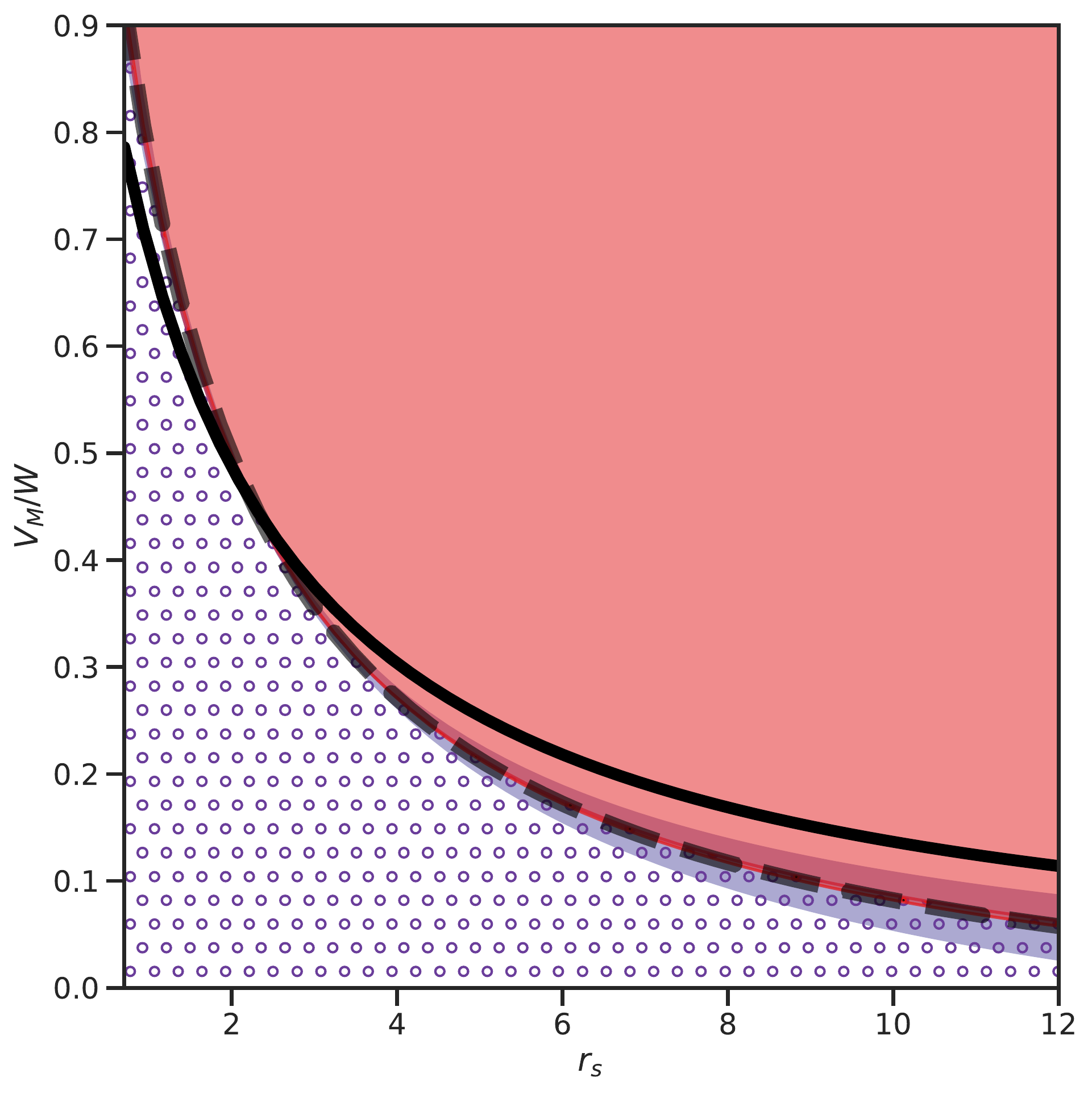}
(c) hybrid LDA
\end{minipage}
\caption{
Qualitatively different phase diagrams created
by various approximations of correlation effects.
Predictions from three independent-particle theories and exact diagonalization of small
systems
are benchmarked by the
result in Fig.~\ref{fig:dmc-phases}, with the black solid line marking the MIT phase boundary from QMC.
The dashed line marks the MIT from each independent-particle theory.
\textbf{(a)} HF predicts an early MIT, which is restricted to high density ($r_s<2.5$), and favors a ferromagnetic state as interaction increases. A small region of collinear stripe phase sits between the noncolinear and ferromagnetic phases.
HF results
for the MIT from the present study and from Ref.~\cite{Hu2021} (light green dots) are in good agreement.
Exact diagonalization predictions for the MIT from Ref.~\cite{Morales-Duran2021} are also shown
(dark green dots), which exhibits growing error at larger $r_s$.
\textbf{(b)} LDA favors the metallic state, thus a late MIT in deep moir\'e potential. It predicts a bandgap closure within the magnetic phase, resulting in a metallic spin density wave (SDW) phase that retains the long-range magnetic order (red dot-hatched region).
The shaded areas around the DFT transition lines are uncertainty estimates due to convergence errors.
\textbf{(c)} Hybrid LDA with $50$\% exact exchange is able to
yield a phase diagram in this system which is in qualitative agreement with QMC.
}
\label{fig:mf-phases}
\end{figure*}

\ssec{Results and Discussions}
Figure~\ref{fig:dmc-phases} shows the results of our QMC phase diagram of the MCH.
As the electron-electron interaction, $U$, and the external moir\'e potential, $V_M$, are increased from zero, the system undergoes a first-order transition from a paramagnetic metal to a $120^\circ$ N\'eel insulator.
The strength of the moir\'e potential $\lambda$ required to induce the transition
decreases monotonically with increasing $r_s$.

The limiting behaviors of the MCH phase diagram are independent of model details.
At constant $V_M$ and in the high-density limit ($r_s\rightarrow0$), we expect a paramagnetic metal because the kinetic energy dominates.
At $V_M=0$, the MCH reduces to the 2DEG, thus we expect a transition from the paramagnetic metal to a Wigner crystal (WC) at $r_s=31\pm1$~\cite{Drummond2009}.
While not explored in this work, we anticipate important changes in the charge and spin properties in the vicinity of the transition as $r_s$ increases towards the WC limit.
For example, magnetic interactions become nearly degenerate in the low-density limit, opening the possibilities for exotic spin states~\cite{Bernu2001}.
While the WC is translationally invariant, any finite $V_M$ pins the WC, allowing its density correlations to be visualized in single-particle densities.

In Fig.~\ref{fig:sdens}, we quantify the spin and charge densities of the metallic and insulating phases.
Density is normalized such that $\int_\Omega \rho(\bs{r})d^2\bs{r}=N$.
Both phases show charge accumulation at the moir\'e minima (A sites) and depletion at the maxima (B sites), whereas only the insulating phase shows significant charge depletion at the saddle points (C sites).
The paramagnetic metal has nearly uniform charge density, with moderate charge accumulation and depletion, peak-to-trough ratio $\sim 2$, which mirror the moir\'e potential.
In contrast, this ratio is $> 15$ in the $120^\circ$ N\'eel phase, where there is little charge at the maxima of the moir\'e potential (B sites).
Site-integrated spin densities, shown as red arrows in the top right panel of Fig.~\ref{fig:sdens}, realize the $120^\circ$ N\'eel magnetic order.
The charge densities have $C_{3z}$ symmetry due to the internal structure of the moir\'e potential at $\phi=26^\circ$, which makes the B and C sites different.
They become equivalent when $\phi=60^\circ$, which could be realized in honeycomb moir\'e materials~\cite{Angeli2021}.

We also compute the
electronic momentum distributions, which are shown in Fig.~\ref{fig:nofk}.
The paramagnetic metal phase has nearly identical momentum distribution
to
the 2DEG dispite the significant amount of external moir\'e potential ($V_M/W=0.2$) imposed upon it.
The Fermi surface remains isotropic at $k_F=\sqrt{2}/r_s$ while the moir\'e potential and electron interaction scatter a small amount of momentum density from inside the Fermi surface to the high-momentum tail.
The secondary Fermi surfaces,
too faint to be visible in the main contour plot,
are noticeable in the linecut around $1.75k_F$ and in the difference contour in the inset.
The $120^\circ$ N\'eel insulator has a smooth momentum distribution with no sign of discontinuity.

Accurate treatment of interaction and correlation is crucial in determining the phase diagram of the MCH.
Our QMC phase diagram (Fig.~\ref{fig:dmc-phases}) is a major revision of that from HF (Fig.~\ref{fig:mf-phases}(a)), where correlation effects are ignored.
A previous exact diagonalization (ED) study~\cite{Morales-Duran2021} found a continuous/weakly first-order MIT, which lies between the HF and QMC predictions.
The small system and basis sizes used in the ED study lead to an underestimation of the gap.
The $120^\circ$ phase has an indirect band gap, so a continuous metal-insulator transition is possible in principle.
However, we find a first-order transition in our most accurate calculations within the finite resolution of
our scan of the two parameters $\lambda$ and $r_s$.
See Supplemental Material [url].
As shown in Fig.~\ref{fig:mf-phases}(b), LDA predicts an early gap closure in the magnetic state, leading to a spin density wave (SDW) phase between the metal and the $120^\circ$ N\'eel insulator.
However, by introducing exact exchange interaction from HF into LDA via a hybrid functional, the charge gap increases to eliminate the SDW phase, making the hybrid LDA phase diagram in qualitative agreement with QMC as shown in Fig.~\ref{fig:mf-phases}(c).
The magnetization disappears abruptly across the transition boundary, driving the charge gap to zero discontinuously.
This hybrid functional can thus potentially serve as an inexpensive
tool for quick first theoretical explorations in these systems,
although it is important to keep in mind its empirical nature,
especially in predicting properties.

All our calculations are at $T=0K$.
To better connect with experiments, we estimate the exchange energy scale by computing the energy cost, $\Delta E$,
to flip a spin in the AFM ``stripe'' phase, which contains alternating stripes of up and down spins.
In a nearest-neighbor Heisenberg model $H=\sum_{\langle i, j\rangle} J S_i S_j$, $\Delta E=4J$.
At $r_s=8$ and $\lambda=1$, we obtain $J \approx 40$ mK,
which is seen to decrease rapidly with increasing $\lambda$.
See Supplemental Material [url] for details.
This is consistent with our total energy comparisons which indicate that the ``stripe'' phase is nearly degenerate with the $120^\circ$ phase.
The near-degeneracy of magnetic states can lead to exotic spin physics, which we hope to explore in future work.

\ssec{Conclusion and Outlook}
We
have characterized the ground-state properties of the MCH
at half-filling in the intermediate to high density regime.
This model captures key ingredients in TMDC systems, namely the presence
of the moir\'e potential and
strong correlation physics of the two-dimensional electron gas,
and
can serve as a fundamental model for providing quantitative understanding of these fascinating systems.
Combining two different QMC methods, we obtain benchmark-quality data on the energetics,
momentum distributions, and the strengths of the magnetic
and charge ordering. The system transitions from a
paramagnetic metal to
a $120^\circ$ N\'eel insulator phase as the moir\'e is deepened or as the density is lowered.
Existing approximate treatments, either via independent-electron approaches or by
simplified lattice models, were seen to result in significant discrepancies in the predicted properties. We tested 2D LDA and hybrid functionals and found that a
hybrid mix of $50$\%
yields a reasonable ground-state phase diagram
in this system
when compared to our
QMC predictions.

We hope that this study paves the way for QMC and other many-body studies of the MCH and
related systems. Many questions remain to be explored, including the physics ---
and potentially more interesting/exotic phases --- at lower
density, with other filling fractions, other structures (moir\'e patterns),
the effect
of spin-orbit coupling and valley degrees of freedom.

\ssec{Acknowledgment}
We thank the Flatiron Institute Scientific Computing Center for computational resources and technical support. The Flatiron Institute is a division of the Simons Foundation.
We thank Andrew Millis, Cody Melton, David Ceperley, Daniele Guerci, Jiawei Zang, Liang Fu, and Scott Jensen for useful discussions.

\bibliographystyle{apsrev4-1}
\bibliography{ref}

\onecolumngrid
\section{Supplementary Material}\label{supplemental}

\begin{figure}[h]
\begin{minipage}{0.48\textwidth}
\includegraphics[width=\linewidth]{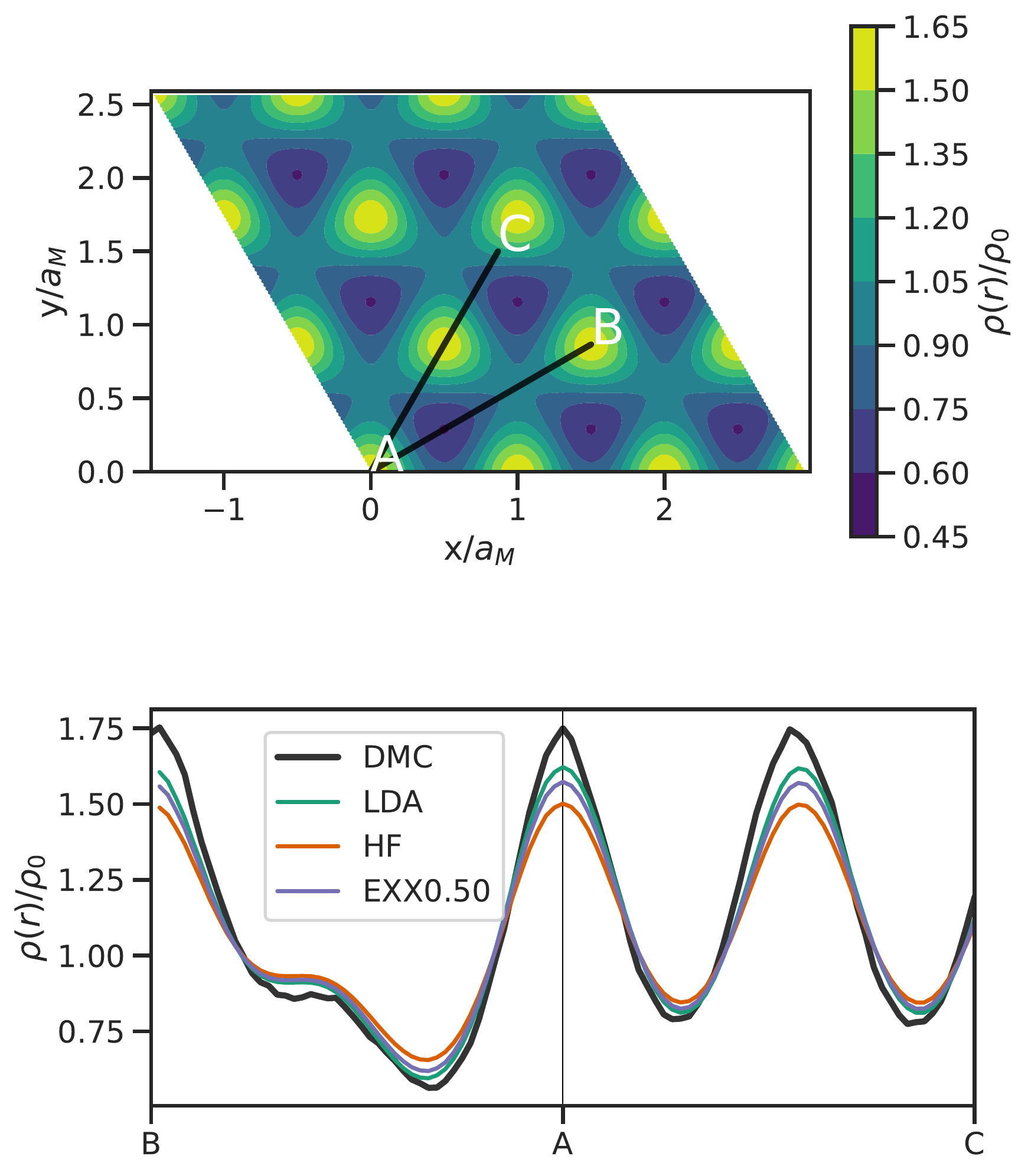}
(a) metal $V_M/W=0.3$
\end{minipage}
\begin{minipage}{0.48\textwidth}
\includegraphics[width=\linewidth]{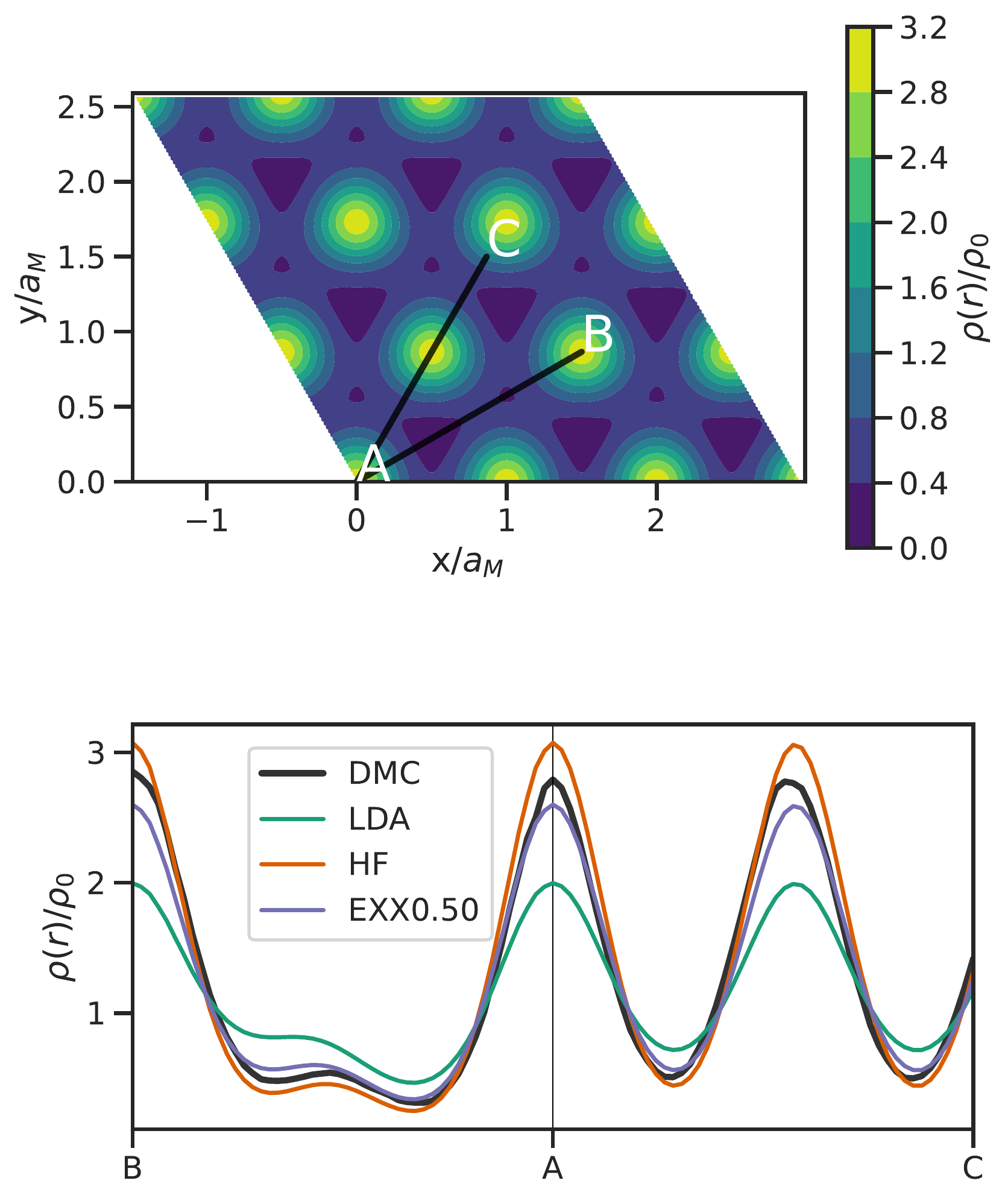}
(b) insulator $V_M/W=0.4$
\end{minipage}
\caption{FP-DMC (N=144) charge density at $r_s=3$ along the linecut BAC. Charge densities from three effective one-body theories are shown along with the QMC linecuts. All functionals produce reasonable charge density for the metallic phase, but only hybrid LDA agrees well with QMC in the insulating phase.}
\label{fig:qmc-dft-cdens}
\end{figure}

\begin{figure}[h]
\begin{minipage}{0.48\textwidth}
\includegraphics[width=\linewidth]{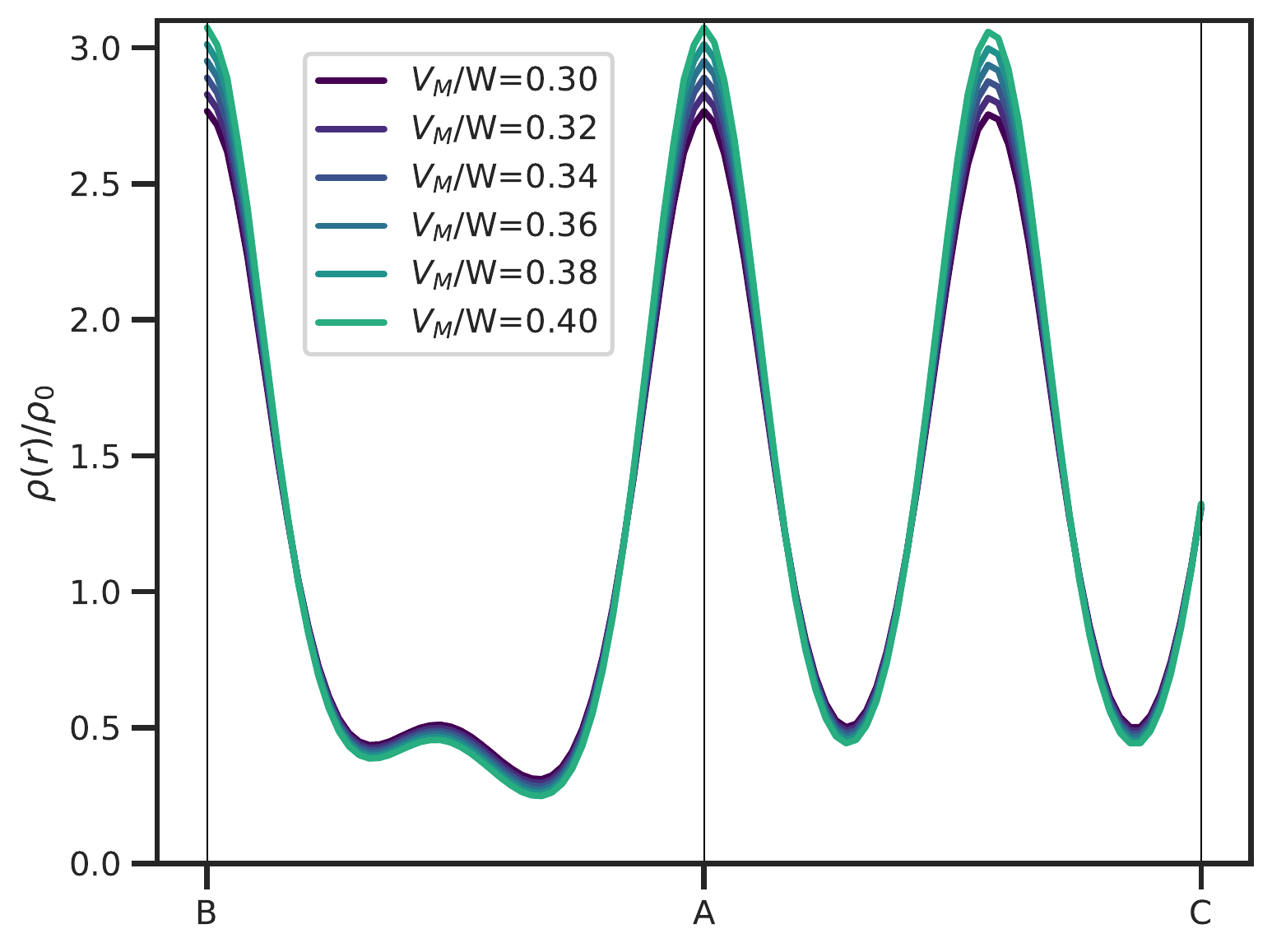}
(a) HF
\end{minipage}
\begin{minipage}{0.48\textwidth}
\includegraphics[width=\linewidth]{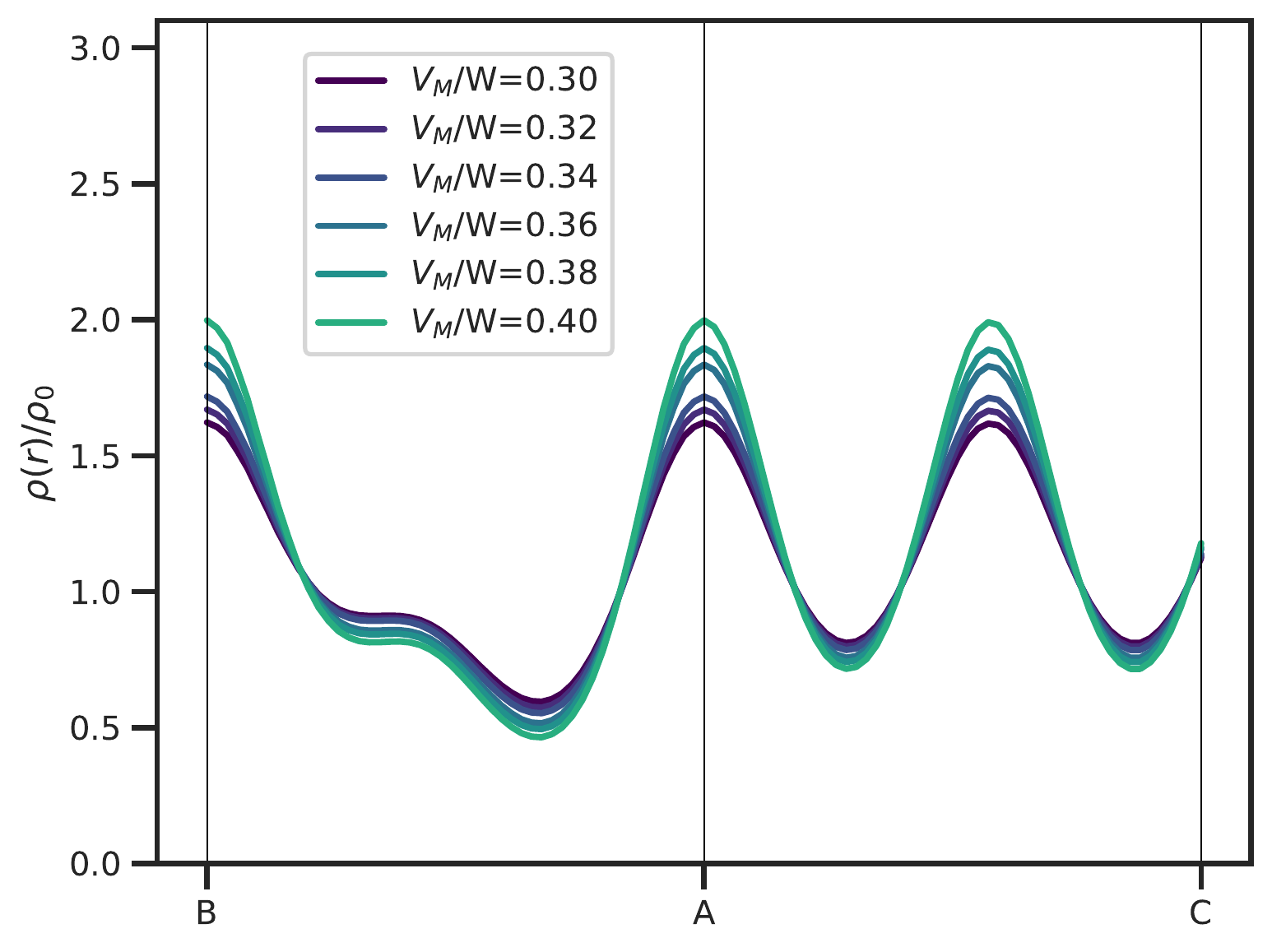}
(b) LDA
\end{minipage}
\begin{minipage}{0.48\textwidth}
\includegraphics[width=\linewidth]{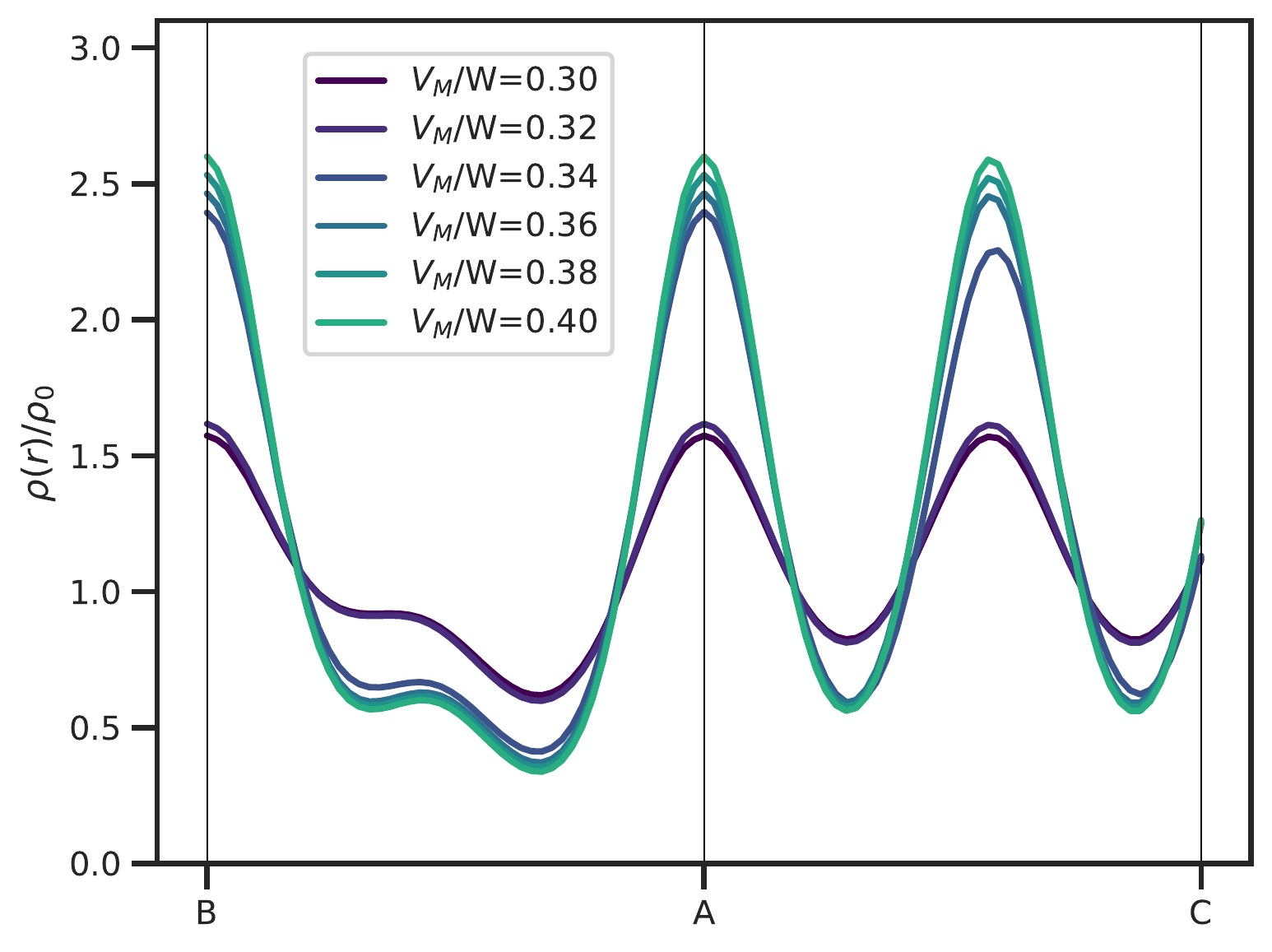}
(c) LDA+$50$\%EXX
\end{minipage}
\begin{minipage}{0.48\textwidth}
\includegraphics[width=\linewidth]{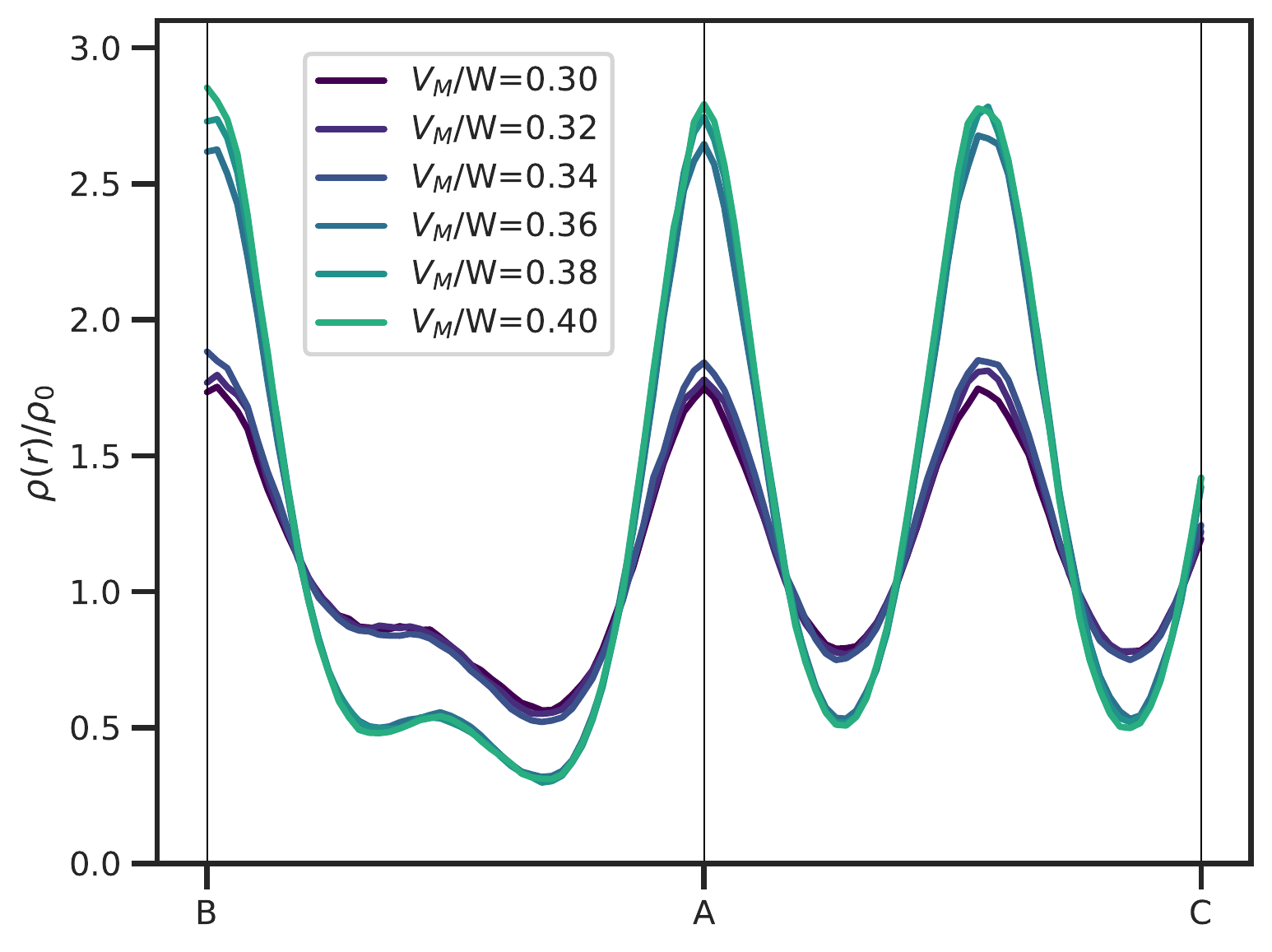}
(d) QMC
\end{minipage}
\caption{Charge density linecuts (same as Fig.~\ref{fig:qmc-dft-cdens}) across the metal insulator transition at $r_s=3$.}
\label{fig:dft-mit}
\end{figure}

Figure~\ref{fig:qmc-dft-cdens} shows FP-DMC, HF, and DFT charge densities before and after the metal-insulator transition (MIT) at $r_s=3$. Hybrid LDA is the only density functional that reasonably reproduces QMC charge density on both sides of the MIT. In Fig.~\ref{fig:dft-mit}, we show the evolution of charge density across the MIT. Only hybrid LDA and QMC have a discontinuous jump across the transition.

Table~\ref{tab:dmc-energy} lists the FP-DMC energy for the calculations shown in Fig.~\ref{fig:qmc-dft-cdens}. At each point on the phase diagram, we compute two FP-DMC energies using Slater-Jastrow trial wavefunctions built from HF and LDA orbitals. We choose the lowest-energy trial to calculate QMC properties. The LDA trial changes from a magnetic insulator to a magnetic metal (spin density wave phase) then to a non-magnetic metal. The HF trial is magnetic and insulating and its FP-DMC energy becomes higher than that of the LDA trial in only the non-magnetic metal phase.

\begin{table}[h]
\caption{Twist averaged FP-DMC energies in Hartree at $r_s=3$.}
\label{tab:dmc-energy}
\begin{minipage}{0.48\textwidth}
(a) metallic phase \\ 
\begin{tabular}{lrrrll}
\toprule
   mag &  $r_s$ &  $V_M/W$ &   N & func &         E/N \\
\midrule
   120 &    3.0 &      0.3 & 144 &   HF & -0.21824(6) \\
   120 &    3.0 &      0.3 & 144 &  LDA & -0.22130(1) \\
  para &    3.0 &      0.3 & 144 &   HF & -0.22121(1) \\
  para &    3.0 &      0.3 & 144 &  LDA & -0.22129(1) \\
stripe &    3.0 &      0.3 & 144 &   HF & -0.21828(5) \\
stripe &    3.0 &      0.3 & 144 &  LDA & -0.22128(1) \\
\bottomrule
\end{tabular}
\end{minipage}
\begin{minipage}{0.48\textwidth}
(b) insulating phase \\
\begin{tabular}{lrrrll}
\toprule
   mag &  $r_s$ &  $V_M/W$ &   N & func &         E/N \\
\midrule
   120 &    3.0 &      0.4 & 144 &   HF & -0.23262(4) \\
   120 &    3.0 &      0.4 & 144 &  LDA & -0.23003(2) \\
  para &    3.0 &      0.4 & 144 &   HF & -0.23022(2) \\
  para &    3.0 &      0.4 & 144 &  LDA & -0.23037(2) \\
stripe &    3.0 &      0.4 & 144 &   HF & -0.23240(4) \\
stripe &    3.0 &      0.4 & 144 &  LDA & -0.23039(2) \\
\bottomrule
\end{tabular}
\end{minipage}
\end{table}

\begin{figure}[h]
\begin{minipage}{0.48\textwidth}
\includegraphics[width=\linewidth]{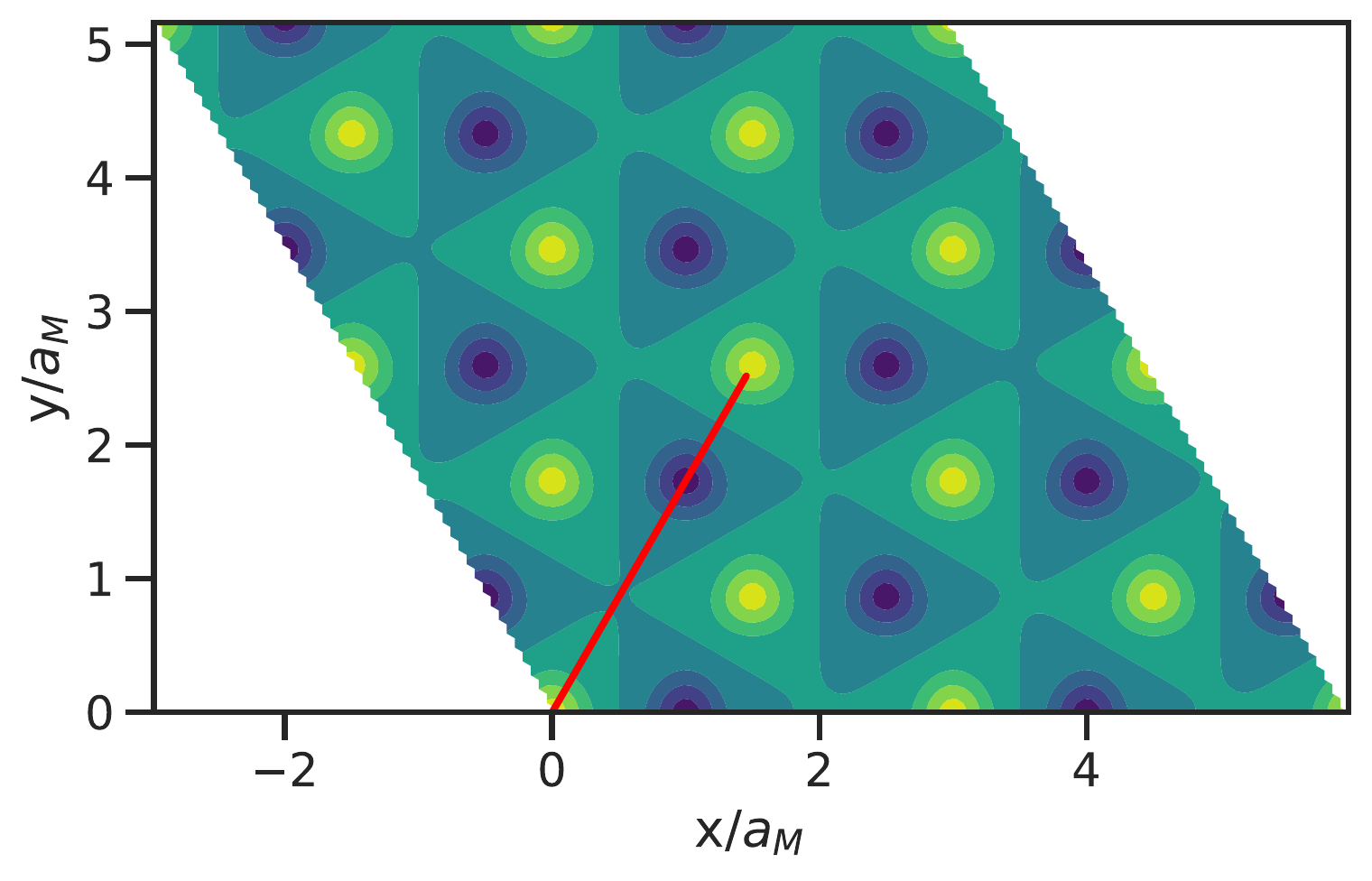}
(a) $\rho_x(x, y)$
\end{minipage}
\begin{minipage}{0.48\textwidth}
\includegraphics[width=\linewidth]{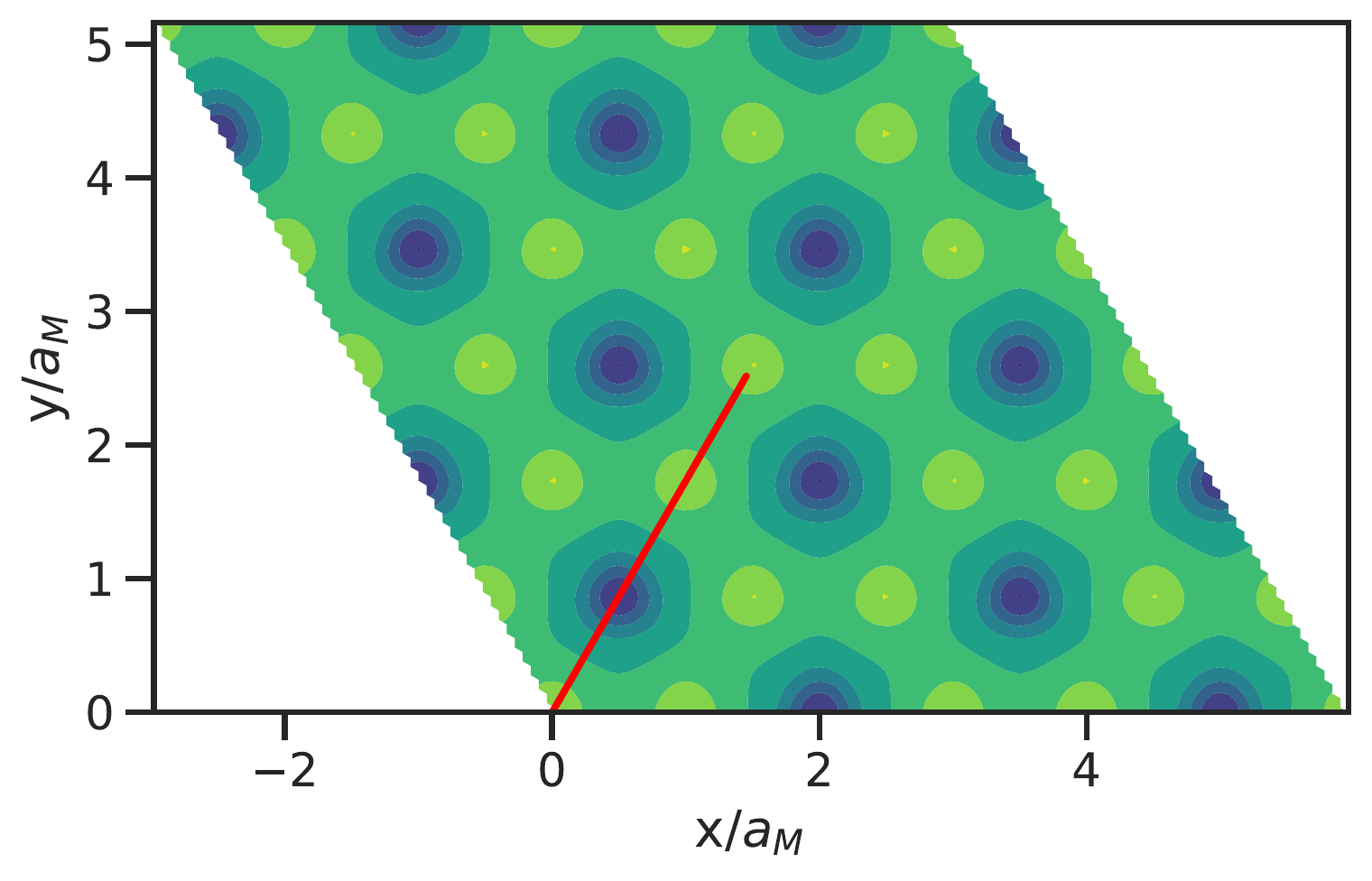}
(b) $\rho_y(x, y)$
\end{minipage}
\begin{minipage}{0.48\textwidth}
\includegraphics[width=\linewidth]{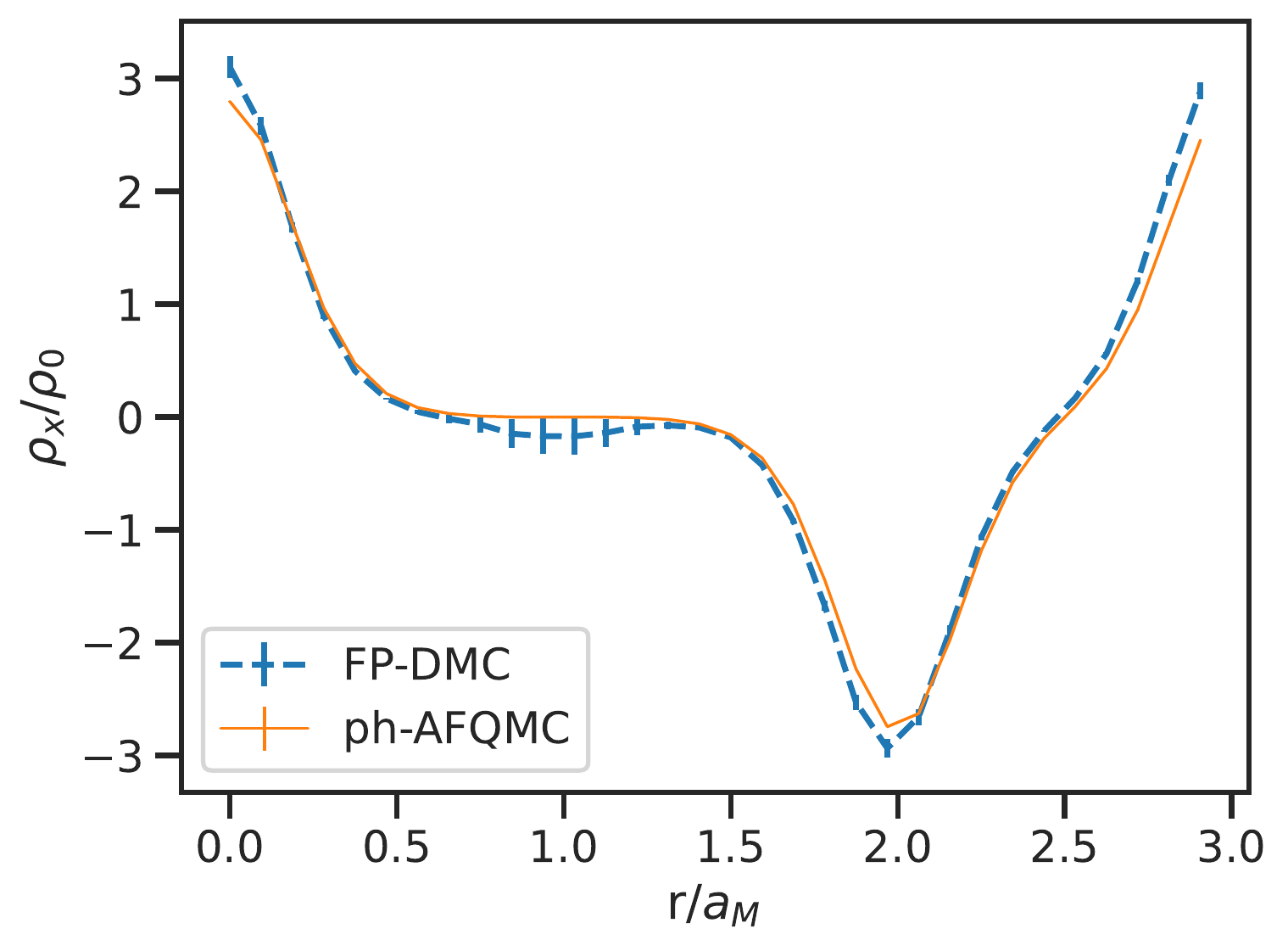}
(c) $\rho_x(r, r)$
\end{minipage}
\begin{minipage}{0.48\textwidth}
\includegraphics[width=\linewidth]{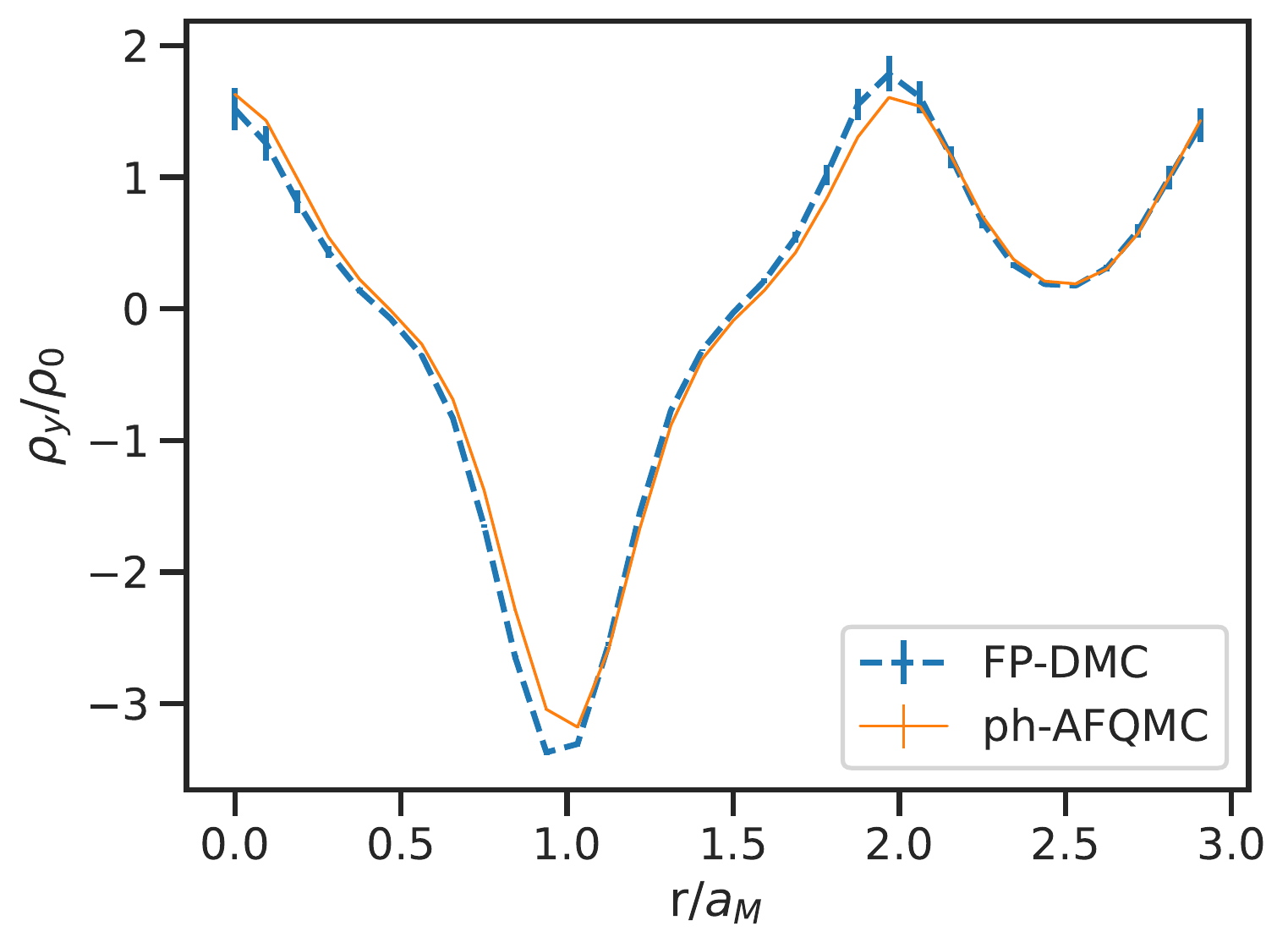}
(d) $\rho_y(r, r)$
\end{minipage}
\caption{Magnetization density of AFQMC and DMC at $r_s=3$ and $\lambda=0.6$.
}
\label{fig:daf-dens}
\end{figure}

Figure~\ref{fig:daf-dens} shows a comparison of FP-DMC and ph-AFQMC magnetization densities of the $120^\circ$ N\'eel phase at $r_s=3$ and $\lambda=0.6$.
Both calculations averaged over $16$ twists, each containing $36$ electrons. On average, $140$ HF orbitals at each twist were used as basis functions in AFQMC.
The DMC calculation of the same system used a Slater-Jastrow trial wavefunction built from the occupied HF orbitals.
Densities from the two methods are consistent despite the very different approximations in the algorithms.

\begin{figure}[h]
\begin{minipage}{0.48\textwidth}
\includegraphics[width=\linewidth]{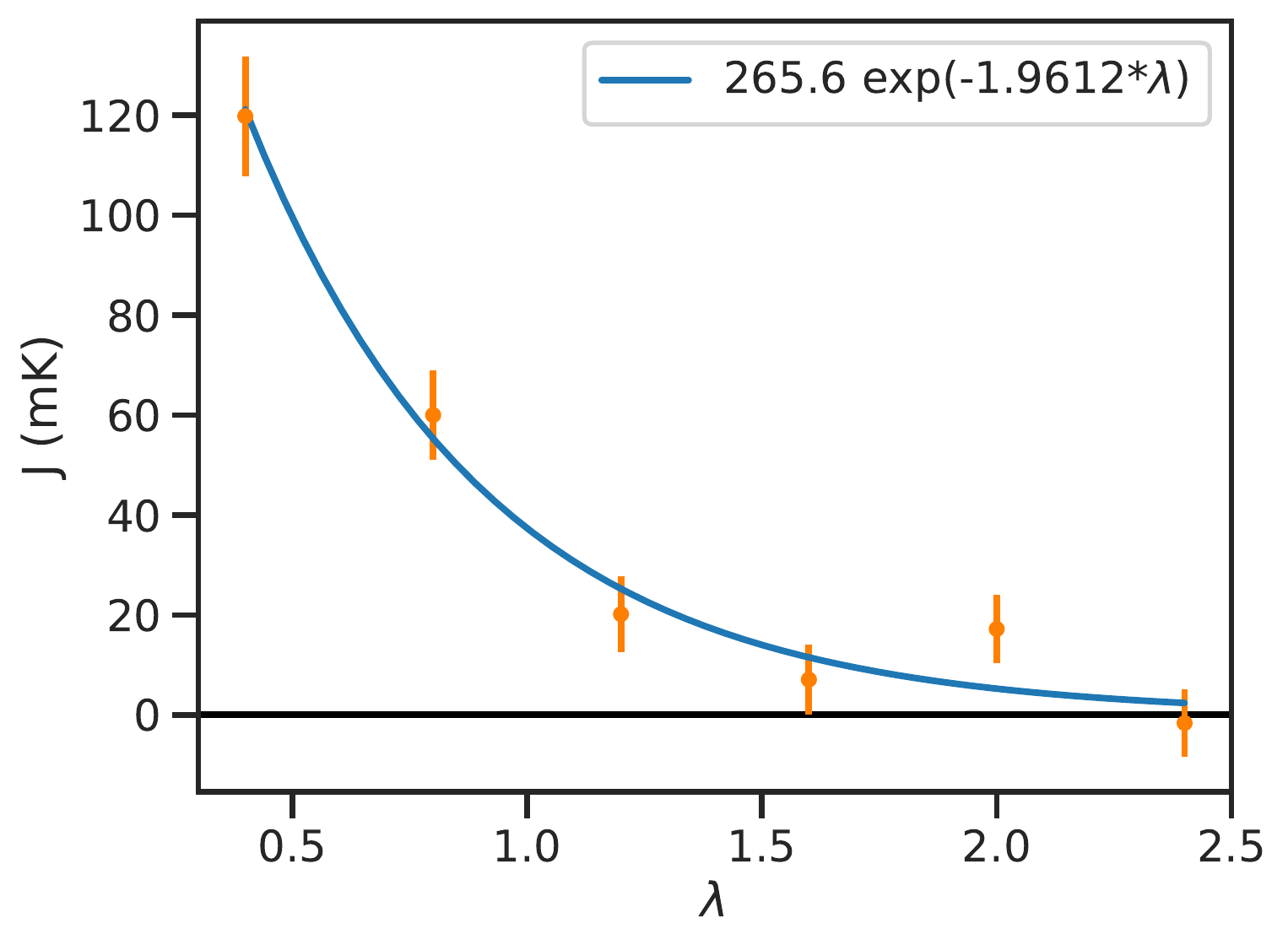}
\end{minipage}
\caption{Estimates of the exchange parameter $J$ of a nearest-neighbor Heisenberg model at $r_s=8$.
}
\label{fig:jheis}
\end{figure}

To make an order-of-magnitute estimate for the magnetic ordering temperature, we calculate the energy cost of a single spin flip $\Delta E$ and related it to the exchange parameter $J=\Delta E/4$ in a nearest-neighbor Heisenberg model of the spin stripe phase.
We made the estimates at $r_s=8$ using a $36$-electron system and FP-DMC with Slater-Jastrow wavefunction and LDA orbitals.
As shown in Fig.~\ref{fig:jheis}, $J$ decreases rapidly from $O(100)$ mK at $\lambda=0.4$ to within errorbars of zero at $\lambda=1.6$.
Given the deep moir\'e potential measured in experiments~\cite{Shabani2021}, we do not expect the ground-state magnetic order to be measurable because of large thermal fluctuations.

\begin{figure}[h]
\begin{minipage}{0.48\textwidth}
\includegraphics[width=\linewidth]{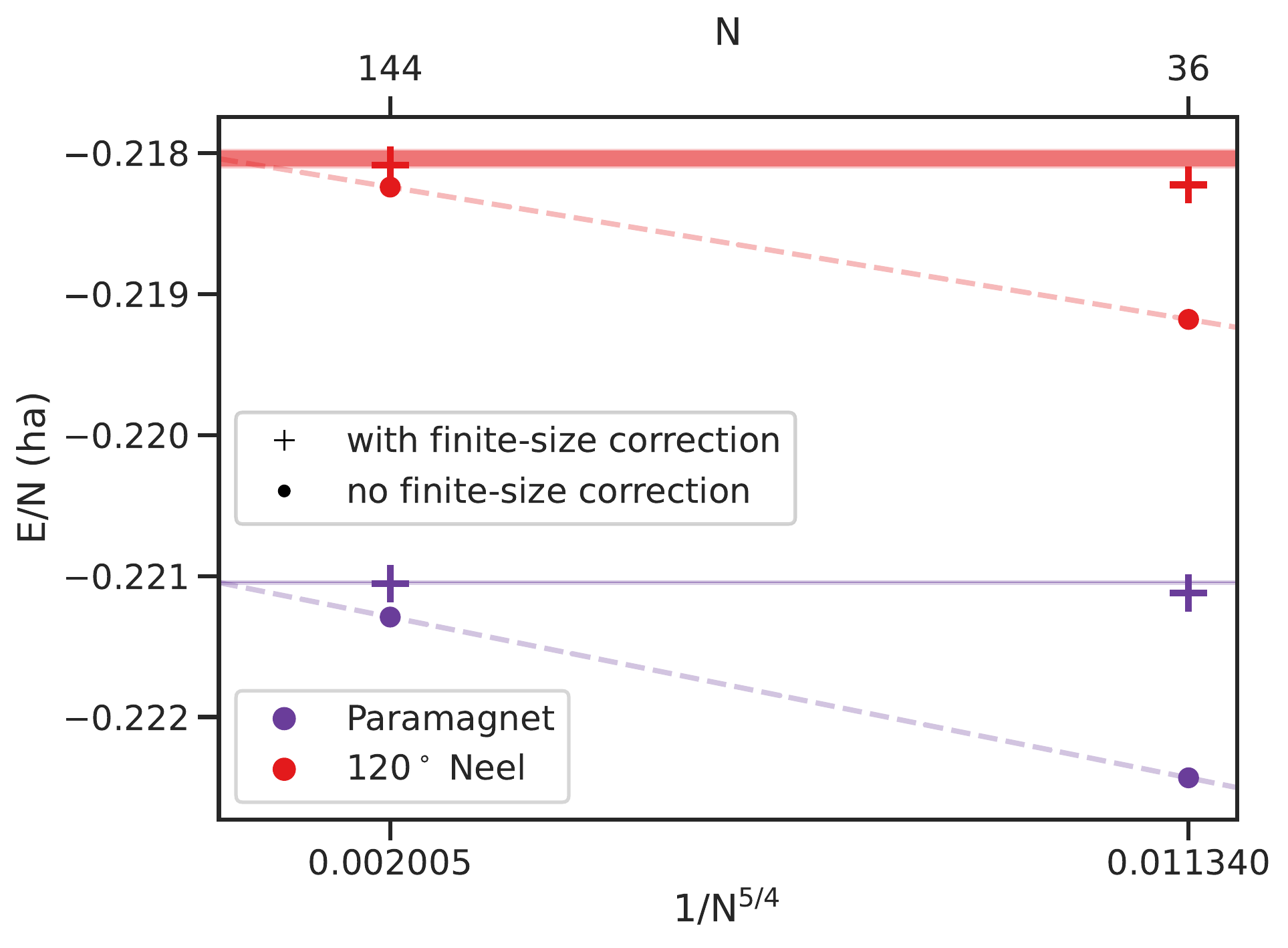}
(a) $V_M/W=0.3$
\end{minipage}
\begin{minipage}{0.48\textwidth}
\includegraphics[width=\linewidth]{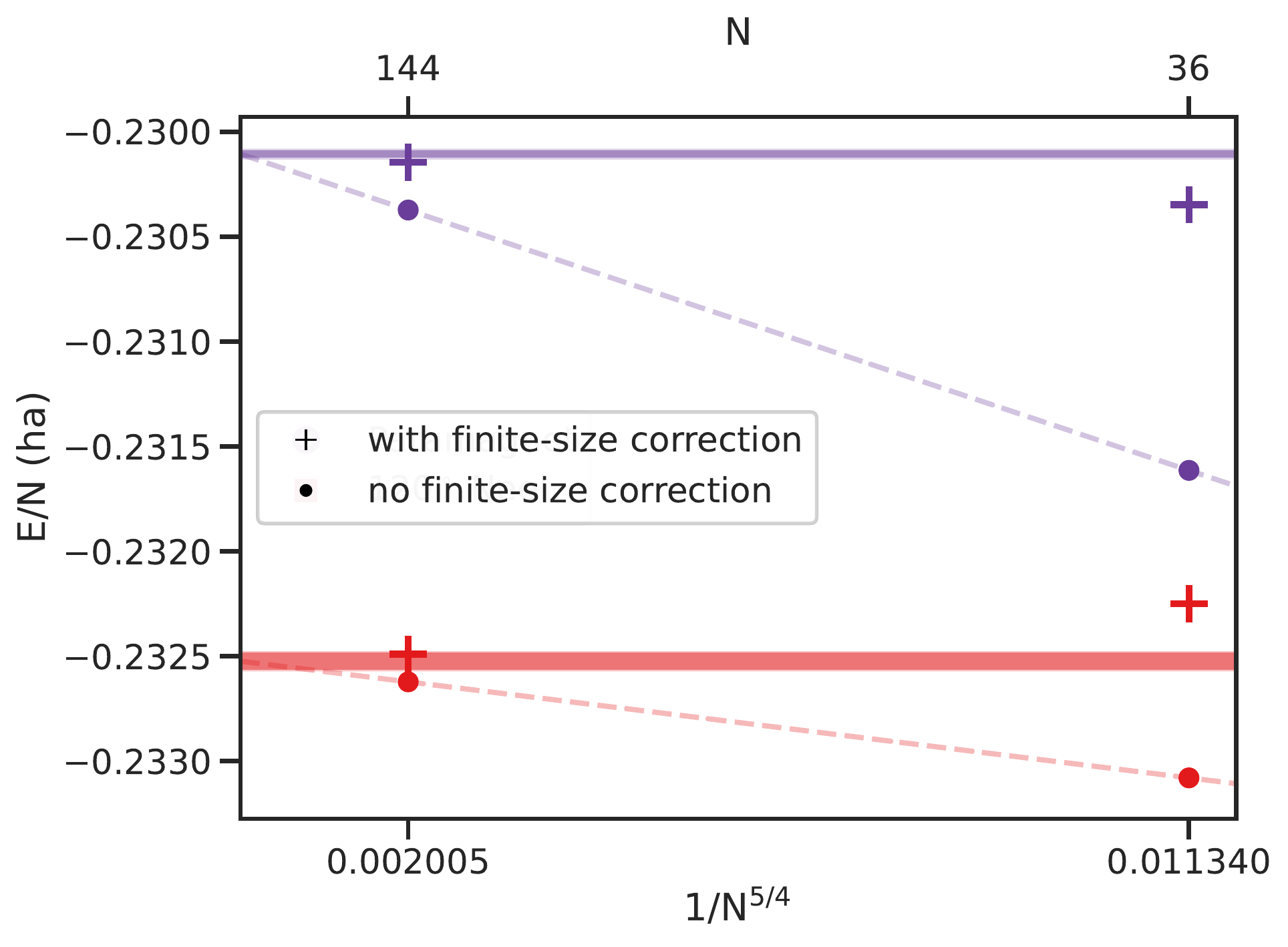}
(b) $V_M/W=0.4$
\end{minipage}
\caption{Finite-size correction (FSC) and extrapolation of FP-DMC total energy at $r_s=3$.
The faint dashed line shows size extrapolation.
The horizontal line and its width represent the mean and error of the thermodynamic limit.
It agrees with $N=144$ result with FSC.
}
\label{fig:nc-extrap}
\end{figure}

\begin{figure}[h]
\begin{minipage}{0.48\textwidth}
\includegraphics[width=\linewidth]{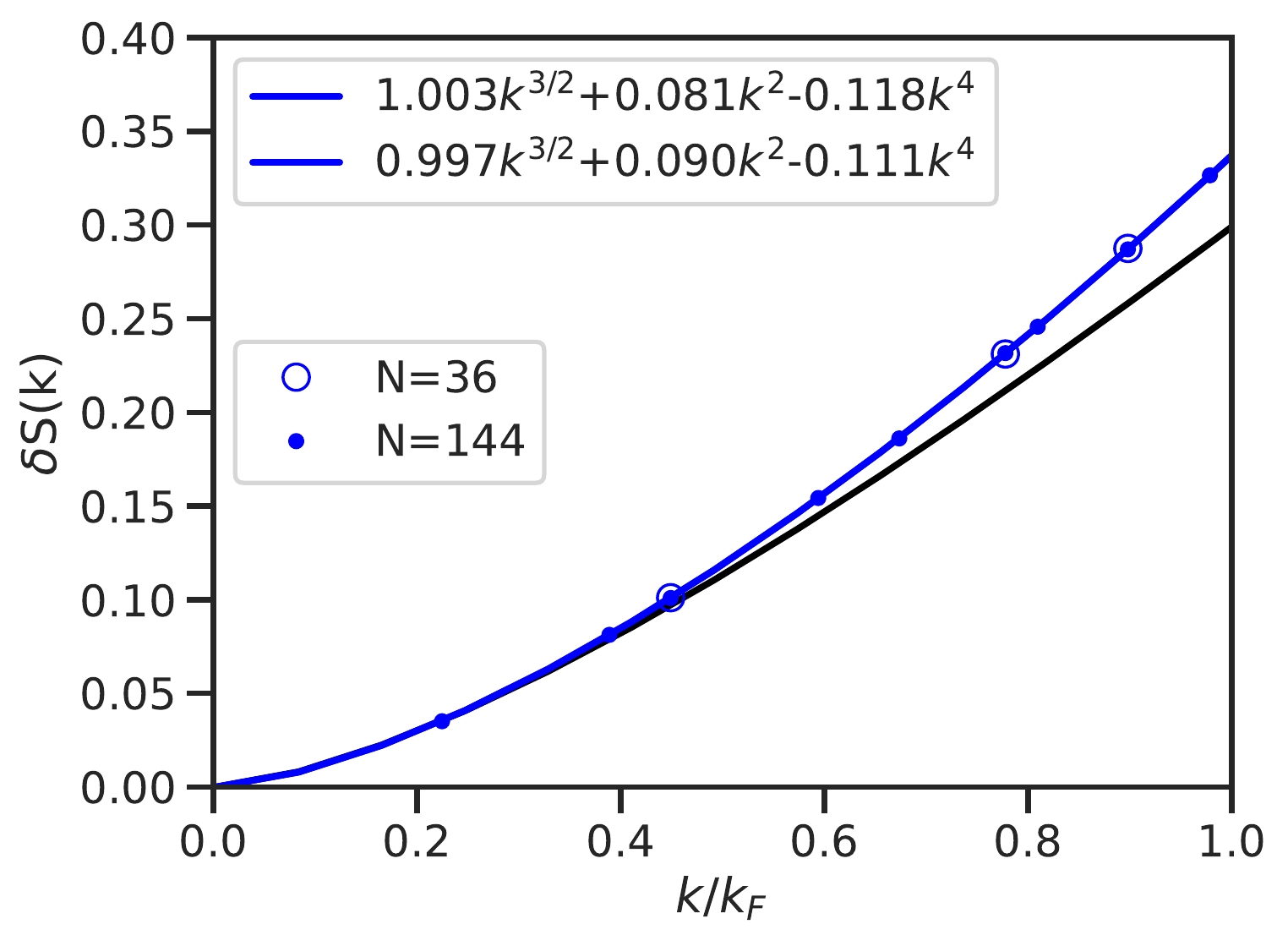}
(a) non-magnetic LDA orbitals $V_M/W=0.3$
\end{minipage}
\begin{minipage}{0.48\textwidth}
\includegraphics[width=\linewidth]{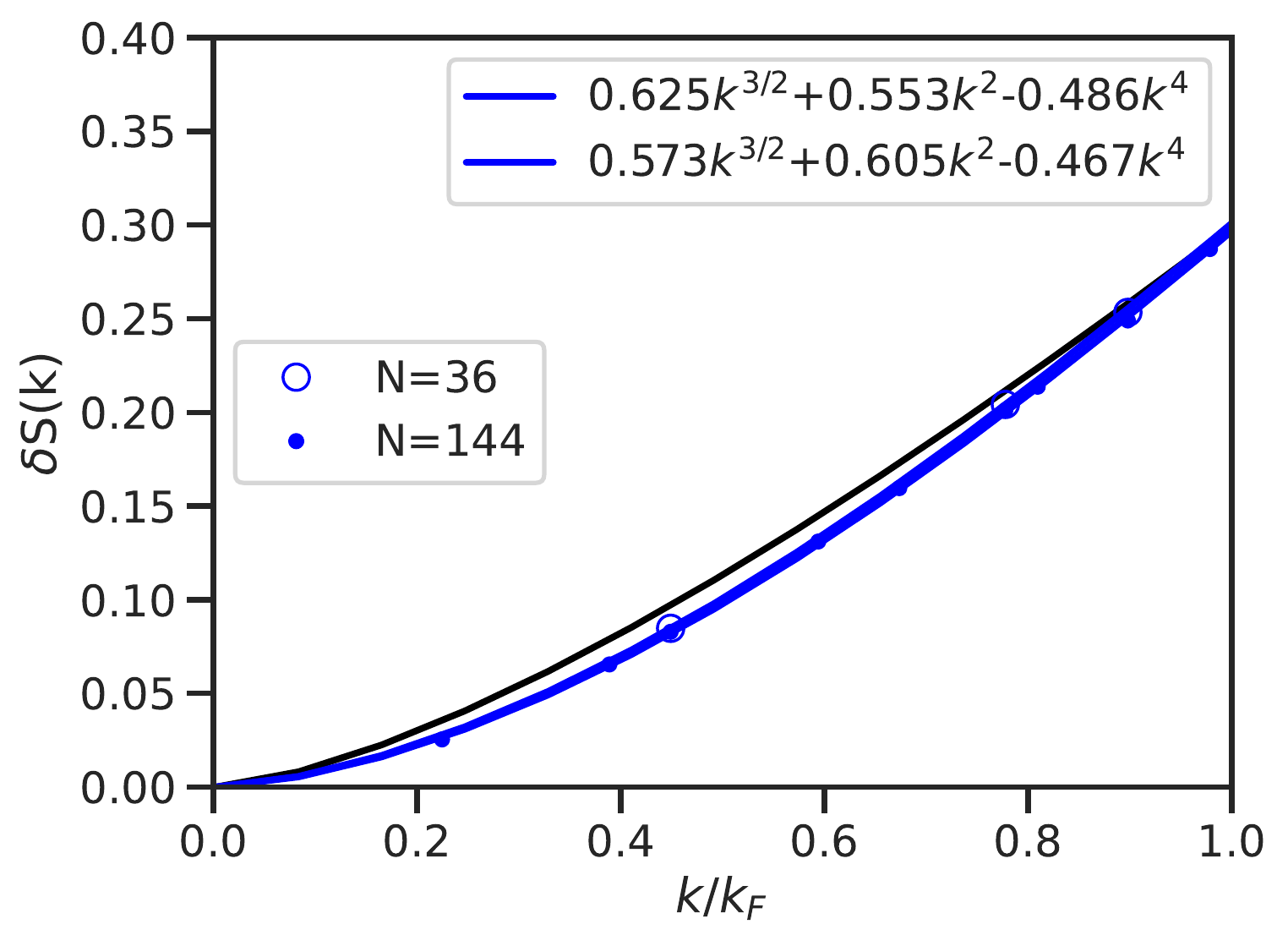}
(b) $120^\circ$ N\'eel HF orbitals $V_M/W=0.3$
\end{minipage}
\begin{minipage}{0.48\textwidth}
\includegraphics[width=\linewidth]{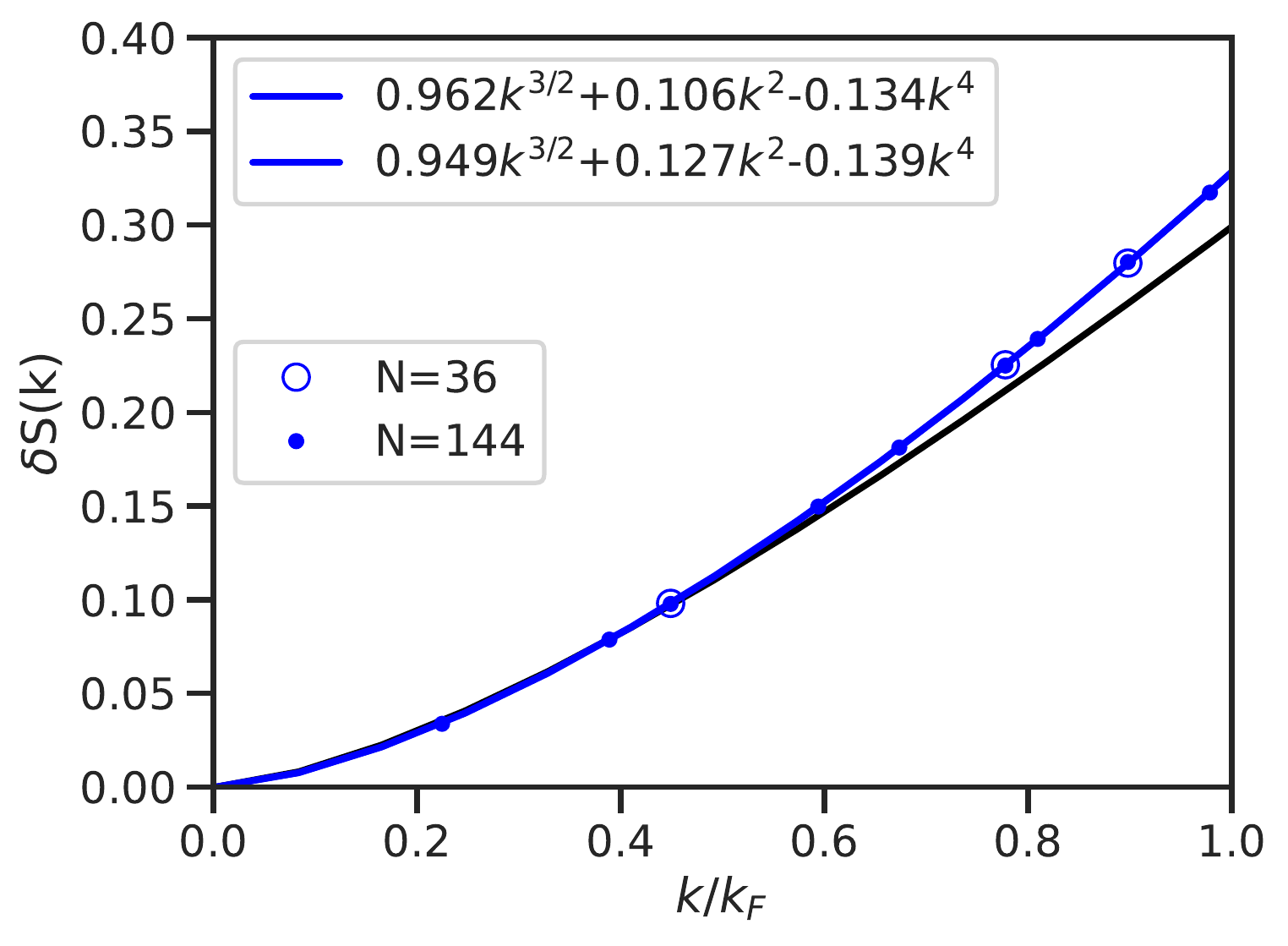}
(c) non-magnetic LDA orbitals $V_M/W=0.4$
\end{minipage}
\begin{minipage}{0.48\textwidth}
\includegraphics[width=\linewidth]{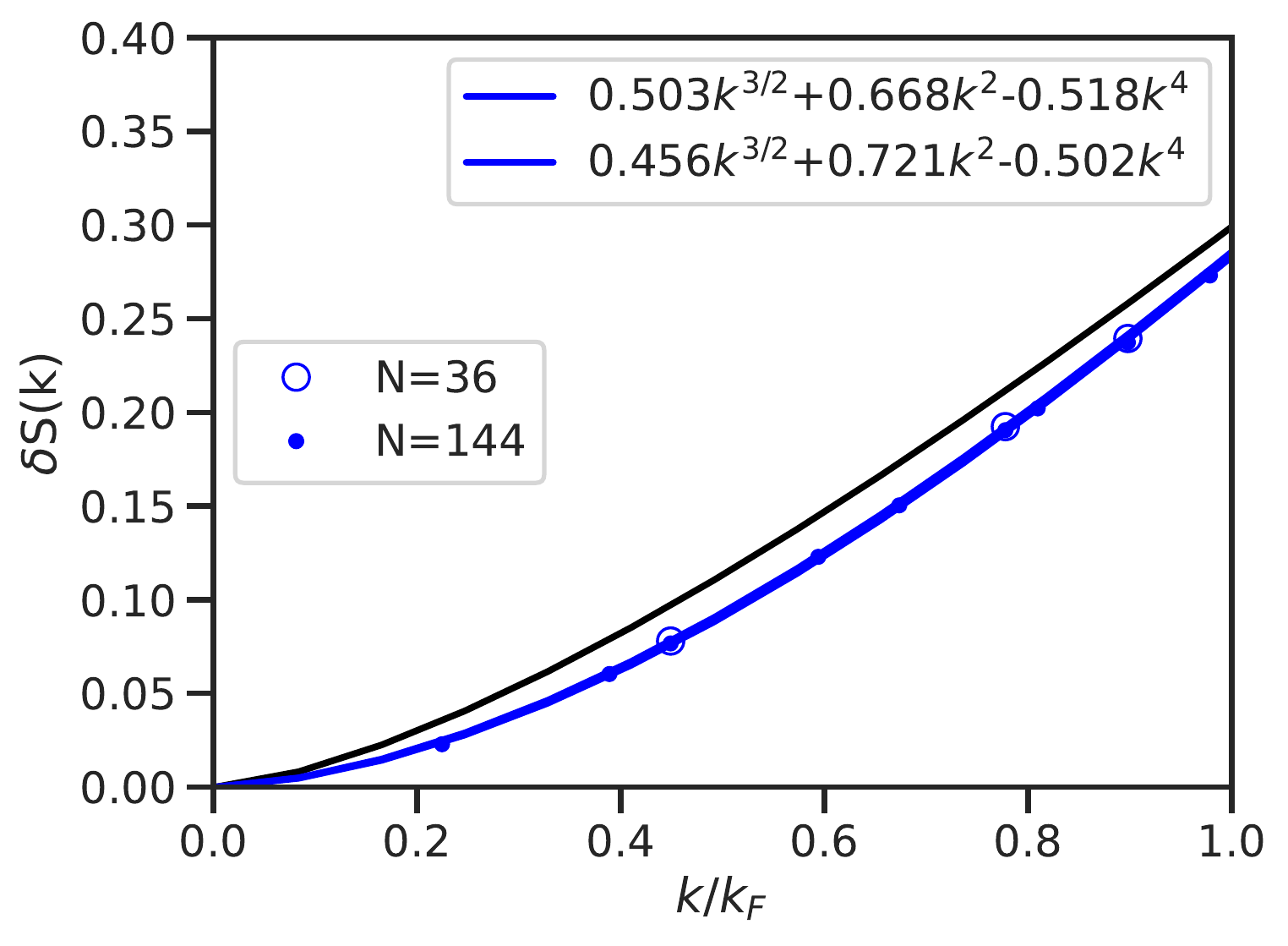}
(d) $120^\circ$ N\'eel HF orbitals $V_M/W=0.4$
\end{minipage}
\caption{Fluctuating structure factor $\delta S(k)$ at $r_s=3$.
The black solid line is reference from random phase approximation (RPA), which exactly captures the long-wavelength plasmon fluctuations in the metallic phase.
Blue lines are polynomial fits to the FP-DMC data.
}
\label{fig:nc-dsk}
\end{figure}

Figure~\ref{fig:nc-extrap} shows finite-size correction (FSC)~\cite{Chiesa2006,Holzmann2016-fsc} and thermodynamic extrapolation~\cite{Drummond2009} at $r_s=3$ and $\lambda=0.3, 0.4$.
The two approaches agree in simulation cells containing $144$ electrons in both the metallic and the insulating phases.
To calculate the FSC, we fit the fluctuating structure factor $\delta S(k)=\langle (\rho_{\bs{k}}-\langle\rho_{\bs{k}}\rangle)(\rho_{-\bs{k}}-\langle\rho_{-\bs{k}}\rangle)\rangle$ to $a k^{3/2} + b k^2 + c k^4$ as shown in Fig.~\ref{fig:nc-dsk}.
The potential energy correction is calculated as $\delta V= [\int -\sum] \frac{1}{2}v_k \delta S(k)$, while the kinetic energy correction is approximated to leading-order $\delta T \approx \delta V$.
These results show that the $144$-electron system is large enough to accurately estimate the total energy in the thermodynamic limit.

\end{document}